\documentclass[aps,pra,twocolumn,showpacs,superscriptaddress]{revtex4-1}  % for review and submission
\usepackage{graphicx}  % needed for figures
\usepackage{dcolumn}   % needed for some tables
\usepackage{bm}        % for math
\usepackage{amsmath, amsthm, amssymb}
\usepackage{mathrsfs}
\usepackage{array}
\usepackage[usenames]{color}
\usepackage{soul} % to use \st for strikeout 
\usepackage[normalem]{ulem} % to use \sout for strikeout --> better than \st: it crosses out across a whole paragraph, including equations, etc.

\begin{document}

\title{Spiral magnetic phases on the Kondo Lattice Model: A Hartree-Fock approach}

\author{Natanael de Carvalho Costa}
\email{natanael@if.ufrj.br}
\affiliation{Instituto de F\'\i sica, Universidade Federal do Rio de Janeiro, Rio de Janeiro, RJ - Brazil}%Lines break automatically or can be forced with \\
\affiliation{Departamento de F\'\i sica, Universidade Federal do Piau\'i, Teresina, PI - Brazil.\\}
\author{Jos\'e Pimentel de Lima}%
\affiliation{Departamento de F\'\i sica, Universidade Federal do Piau\'i, Teresina, PI - Brazil.\\}%

\author{Raimundo R. \surname{dos Santos}}
\affiliation{Instituto de F\'\i sica, Universidade Federal do Rio de Janeiro, Rio de Janeiro, RJ - Brazil}%

\date{\today}

\begin{abstract}
We study the Kondo Lattice Model (KLM) on a square lattice through a Hartree-Fock approximation in which the local spins are treated semi-classically, in the sense that their average values are modulated by a magnetic wavevector $\mathbf{Q}$ while they couple with the conduction electrons through fermion operators.
In this way, we obtain a ground state phase diagram in which spiral magnetic phases (in which the wavevector depends on the coupling constants and on the density) interpolate between the low-density ferromagnetic phase and the antiferromagnetic phase at half filling; 
within small regions of the phase diagram commensurate magnetic phases can coexist with Kondo screening.
We have also obtained `Doniach-like' diagrams, showing the effect of temperature on the ground state phases, and established that for some ranges of the model parameters (the exchange coupling and conduction electron density) the magnetic wavevector changes with temperature, either continuously or abruptly (e.g., from spiral to ferromagnetic). 
\end{abstract}

\pacs{
71.27.+a, 	%Strongly correlated electron systems; heavy fermions
71.10.Fd, 	%Lattice fermion models (Hubbard model, etc.)
75.10.-b,	%General theory and models of magnetic ordering (see also 05.50.+q Lattice theory and statistics)
75.30.Mb	%Valence fluctuation, Kondo lattice, and heavy-fermion phenomena (see also 71.27.+a Strongly correlated electron systems, heavy fermions; for heavy-fermion superconductors, see 74.70.Tx)
}
\maketitle

\section{\label{sec1}Introduction}

The Kondo Lattice Model (KLM) \cite{Doniach1977,Lacroix1979} describes a system consisting of magnetic moments occupying each site of a regular lattice, and interacting with conduction electrons through a local exchange interaction, $J$; see, e.g., Refs.\,\onlinecite{Tsunetsugu97,Coleman2007}.
The exchange coupling leads to two types of effects: the Kondo screening and the Ruderman-Kittel-Kasuya-Yosida interaction (RKKY). 
The former creates a hybridization between conduction electrons and local moments, which favours a paramagnetic Fermi liquid phase in which the local moments contribute to the Fermi surface;
the latter creates an indirect interaction between the local moments, through the polarization of the conduction electrons, thus favouring a magnetically ordered state with a small Fermi surface accommodating solely conduction electrons. 
As first discussed by Doniach \cite{Doniach1977}, these two effects can coexist, hence compete with each other, leading to a quantum phase transition from a magnetically ordered state to a paramagnetic one.

This competition between Kondo screening and magnetism, including the existence of a quantum critical point, is known to occur in several  heavy-fermion materials \cite{Coleman2007}; in view of this, it is generally believed that the KLM provides an adequate description of some aspects of these materials. 
Further, early mean-field approaches to the Kondo-lattice Hamiltonian \cite{Lacroix1979,Fazekas1991} already pointed out that for small screening, $J\lesssim W$ ($W$ is the bandwidth), the magnetic ground state is either ferromagnetic or antiferromagnetic, respectively for electronic densities $0\leq n_c\lesssim 0.6$ and $0.6\lesssim n_c \leq 1$.
This immediately connects with an interesting class of materials, the borocarbide family \cite{Muller2001}, which has the chemical composition $R T_{2}$B$_{2}$C, where $R$ represents a rare-earth element and $T$ is a transition metal. 
The rare earth usually contributes with one local magnetic moment per unit cell, which, in turn, interacts with the conduction electrons. 
Some members of the family display coexistence between superconductivity and magnetism \cite{Nagarajan94,Cava94,Schmidt1997,Muller2001,Gupta06}, the interplay of which is a subject of current interest as a unifying link with the iron pnictides \cite{Ishida09}, and, possibly, with the cuprates \cite{Mukuda12,Scalapino12}. 
Setting aside the superconducting behaviour of some borocarbides, one notes that a wide variety of magnetic orderings (or \emph{modes}) have been found: depending on the particular combination of $R$ and $T$, one finds ferromagnetism, antiferromagnetism, as well as spin-density waves, and multiple $\mathbf{Q}$-wavevectors, commensurate and incommensurate \cite{Lynn97,Muller2001}.
This variety of magnetic modes has been further scrutinized with the synthesis of borocarbides with variable proportions of transition metals, as in Tb(Co$_{x}$Ni$_{1-x}$)$_{2}$B$_{2}$C [Refs.\,\onlinecite{ElMassalami2012,ElMassalami2013}] and Ho(Co$_{x}$Ni$_{1-x}$)$_{2}$B$_{2}$C [Ref.\,\onlinecite{ElMassalami2014}].
Notwithstanding the fact that the KLM does not incorporate explicitly some aspects of the borocarbides (such as crystal field effects), one may wonder whether it can be used as an effective model to describe the evolution of magnetic modes with the band filling. 

At any rate, the KLM is an interesting model in its own right, and a great deal of theoretical effort has been invested to unveil its overall properties.
One-dimensional topologies are amenable to unbiased methods such as the Density Matrix Renormalization Group (DMRG), and, indeed, ferromagnetism and spiral magnetic phases can be stabilized in the linear chain, in different regions of the ground state parameter space $(J,n_c)$, where $n_c$ is the density of conduction electrons \cite{Garcia2004};
further, the two-leg version of the model exhibits quasi--long-range magnetic order, with the magnetic wavevector displaying a well defined dependence with the electronic density \cite{Xavier2004}.
In two dimensions, auxiliary-field Quantum Monte Carlo (QMC) results are available at half filling \cite{Assaad99,Capponi01}; away from half filling, variational QMC has been used, but only antiferromagnetic solutions were probed \cite{Watanabe2007,Asadzadeh2013}.
In view of the intrinsic difficulties of those methods to extract comprehensive and simultaneous information (i) about the various magnetic arrangements, (ii) about the interplay between Kondo screening and magnetism, (iii) in two- and three dimensions, (iv) for all conduction electronic densities (from 0 to half filling), and (v) the effects of temperature, many mean-field implementations have been used over the years, tackling some of these issues.   
While several studies only allowed for para-, ferro- and antiferromagnetic phases \cite{Lacroix1979, Fazekas1991, Zheng1996, Zhang2000, Zhang2010, Asadzadeh2013, BenHur2015}, the possibility of $\mathbf{Q}$-dependent magnetic modes was considered in Ref.\,\onlinecite{Hamada1995}, in which case the local spins were treated classically, thus precluding the analysis of the coexistence of Kondo screening with magnetic phases. 
By contrast, this coexistence has so far  been examined only in conjunction with ferromagnetic (FM) or antiferromagnetic (AFM) phases \cite{Zheng1996,Zhang2000,Zhang2010,Asadzadeh2013}. 

Therefore, a mean-field investigation of the KLM taking into account both generic magnetic orderings \emph{and} the effect of Kondo screening is clearly in order. 
With this in mind, here we use a `semi-classical' approach, in which the local moments display a $\mathbf{Q}$-dependent average magnetization, while they are also expressed in terms of fermionic operators allowing us to define a hybridization `order parameter' as a measure of the Kondo screening.
In this quest, we are led to minimize the free energy also with respect to the magnetic wavevector, in order to establish the dependence of the stable $\mathbf{Q}$ values with $J$, $n_c$, and the temperature, $T$. 
Since this is more readily carried out in two dimensions than in three, we choose to consider here the KLM on a square lattice.
Our main results can be summarized  in the form of a phase diagram with many magnetically-ordered phases which evolve into a screening-dominated (Kondo) one, but going through intermediate regions in which they coexist. 

This paper is organized as follows: The model is presented in Sec.\,\ref{sec2}, together with highlights of the standard Hartree-Fock approximation, the details of which can be found in the Appendix. 
In Section \ref{sec3}, we discuss the results for the ground state, while Section \ref{sec4} is devoted to finite temperature behaviour. 
And, finally, Section \ref{sec5} summarizes our findings.

%%%%%%%%%%%%%%%%%%%%%%   MODEL AND METHOD   %%%%%%%%%%%%%%%%%%%%%%%%%%%%%%%%%%%%%%%%

\section{\label{sec2}Model and Method}

The Kondo lattice model is described by the Hamiltonian
\begin{equation} \label{hamiltonian} 
\mathcal{H} = -t\!\sum_{\langle i, j \rangle, \sigma} \big( c^{\dagger}_{i\sigma} c^{\phantom{\dagger}}_{j\sigma} + \mathrm{H.c.} \big) + J \sum_{i} \mathbf{S}_{i}\! \cdot \mathbf{s}^{c}_{i},
\end{equation}
where the sums run over sites of a two-dimensional square lattice, with $\langle i,j \rangle$ denoting nearest-neighbor sites. 
The first term represents the hopping of conduction electrons, where $c^{\dagger}_{i\sigma}$ ($c_{i\sigma}$) is the creation (annihilation) operator for an electron on site $i$ with spin $\sigma$, and H.c.\ stands for hermitian conjugate of the previous expression; $t$ sets the energy scale. 
The second term represents an interaction between local moments and conduction electrons, where $J>0$ is the coupling strength, and $ \mathbf{S}_{i}$ and $\mathbf{s}^{c}_{i}$ are the spin operators for the local moment and conduction electrons, respectively. 

In order to set up a Hartree-Fock approximation, we write the spin operators in a fermionic basis as
\begin{equation} \label{Sf}
\mathbf{S}_{i} = \frac{1}{2}\sum_{\alpha, \beta = \pm } f^{\dagger}_{i \alpha} \boldsymbol{\sigma}_{\alpha, \beta}f^{\phantom{\dagger}}_{i \beta} ,
\end{equation}
and
\begin{align} \label{Sc}
\mathbf{s}^{c}_{i} = \frac{1}{2}\sum_{\alpha, \beta = \pm} c^{\dagger}_{i \alpha} \boldsymbol{\sigma}_{\alpha, \beta} c^{\phantom{\dagger}}_{i \beta} ,
\end{align}
with $ \boldsymbol{\sigma}_{\alpha, \beta}$ denoting Pauli matrix elements, and $f^{\dagger}_{i\sigma}$ ($f_{i\sigma}$) being the creation (annihilation) operator for a localized electron with spin $\sigma$ on site $i$. 
Following the procedure outlined in the Appendix, the Hartree-Fock Hamiltonian becomes
\begin{widetext}
\begin{align} 
%\begin{eqnarray}
\nonumber \mathcal{H}_{MF} = & -t \sum_{\langle i,j \rangle, \sigma} \big( c^{\dagger}_{i \sigma} c^{\phantom{\dagger}}_{j \sigma} + \mathrm{H.c.} \big) 
+ J \sum_{i} \big( \mathbf{S}_{i}\! \cdot\! \langle \mathbf{s}^{c}_{i} \rangle + \langle \mathbf{S}_{i} \rangle \!\cdot\! \mathbf{s}^{c}_{i} \big) 
+ \frac{J}{2} \sum_{i} \big( \mathbf{V}^{c}_{i}\! \cdot\! \langle \mathbf{V}^{f}_{i} \rangle + \langle \mathbf{V}^{c}_{i} \rangle \!\cdot\! \mathbf{V}^{f}_{i} \big) \\ 
&- \frac{3 J}{2} \sum_{i} \big( V^{0}_{i c} \langle V^{0}_{i f} \rangle + \langle V^{0}_{i c} \rangle V^{0}_{i f} \big) - \frac{J}{2} \sum_{i} \langle \mathbf{V}^{c}_{i} \rangle\! \cdot\! \langle \mathbf{V}^{f}_{i} \rangle + \frac{3 J}{2} \sum_{i} \langle V^{0}_{i c} \rangle \langle V^{0}_{i f} \rangle
- J \sum_{i} \langle \mathbf{S}_{i} \rangle\! \cdot\! \langle \mathbf{s}^{c}_{i} \rangle,
\label{hamil2}
%\end{eqnarray}
\end{align}
\end{widetext}
with the definitions
\begin{eqnarray}\label{singlet_hyb}
V^{0}_{i c}= {V^{0}_{i f}}^{\dagger} = \frac{1}{2}\sum_{\alpha, \beta = \pm} c^{\dagger}_{i \alpha} \mathbb{I}_{\alpha, \beta} f^{\phantom{\dagger}}_{i \beta},
\end{eqnarray}
where $\mathbb{I}$ is the identity matrix, and
\begin{eqnarray}\label{triplet_hyb}
\mathbf{V}_{i c} = \mathbf{V}^{\dagger}_{i f} = \frac{1}{2} \sum_{\alpha, \beta = \pm} c^{\dagger}_{i \alpha} \boldsymbol{\sigma}_{\alpha, \beta} f^{\phantom{\dagger}}_{i \beta}.
\end{eqnarray}
Following the nomenclature introduced in Ref.\,\onlinecite{K.S.Beach2007}, we refer to 
$ V^{0}_{i c} $ and $ V^{0}_{i f} $ as \emph{singlet} hybridization operator, and to $\mathbf{V}_{i c}$ and $\mathbf{V}_{i f}$ as \emph{triplet} hybridization operators. 

In order to analyse the stability of planar spiral magnetic phases, the mean value $\langle \mathbf{S}_{i} \rangle$ is taken as classical, 
\begin{align}\label{Si}
\langle \mathbf{S}_{i} \rangle = m_{f}^0  \big[ \cos \left(\mathbf{Q}\!\cdot\! \mathbf{R}_{i}\right), \sin \left(\mathbf{Q}\!\cdot\! \mathbf{R}_{i}\right), 0 \big] ,
\end{align}
with 
\begin{equation} 
\label{Q}
\mathbf{Q}=(q_{x}, q_{y})
\end{equation}
being the magnetic wavevector, and $\mathbf{R}_{i}$ the vector position of site \textit{i} on the lattice. %For simplicity, we define $m_{f} (\mathbf{R}_{i}) = m_{f}^{0} + m_{f}^{1} \cos\left( \mathbf{Q}\!\cdot\! \mathbf{R}_{i} \right) $, which allows the magnetization strength of the $f$-sites to vary in space (modulated by the wavevector $\mathbf{Q}$) around a mean value $m_{f}^{0}$.
By the same token, we choose 
\begin{align}\label{Sci}
\langle \mathbf{s}^{c}_{i} \rangle = -m_{c}^0 \big[ \cos \left(\mathbf{Q}\!\cdot\! \mathbf{R}_{i}\right), \sin \left(\mathbf{Q}\!\cdot\! \mathbf{R}_{i}\right),0 \big],
\end{align}
%with $m_{c} (\mathbf{R}_{i}) = m_{c}^{0} + m_{c}^{1} \cos\left( \mathbf{Q}\!\cdot\! \mathbf{R}_{i} \right) $; 
where the minus sign above reflects the local antiferromagnetic coupling between the local moments and the conduction electrons.

The singlet hybridization terms can be taken as
\begin{equation}\label{singlet_hyb_mean_v}
\langle V^{0}_{i c} \rangle = \langle {V^{0}_{i f}}^{\dagger} \rangle = -V_{0},
\end{equation}
and  the mean values of the triplet hybridization operators are similarly assumed to be given by
\begin{equation}\label{triplet_hyb_mean_v}
\langle \mathbf{V}_{i c} \rangle  = \langle  {\mathbf{V}_{i f}}^{\dagger} \rangle = V_0'
\big[ \cos \left(\mathbf{Q}\!\cdot\! \mathbf{R}_{i}\right), \sin \left(\mathbf{Q}\!\cdot\! \mathbf{R}_{i}\right),0 \big].
\end{equation}
%with $V'(\mathbf{R}_{i})= V_{0}'+ V_{1}'\cos\left(\mathbf{Q}\!\cdot\! \mathbf{R}_{i}\right)$.

The electronic density, $n_{c}$, and the number of local moments per site, respectively expressed by
\begin{equation}
\frac{1}{N} \sum_{i \sigma} c^{\dagger}_{i \sigma} c^{\phantom{\dagger}}_{i \sigma} = n_{c},
\end{equation}
and 
\begin{equation}
\frac{1}{N} \sum_{i \sigma} f^{\dagger}_{i \sigma} f^{\phantom{\dagger}}_{i \sigma} = 1, 
\end{equation}
are imposed as constraints through the method of Lagrange multipliers. 
The latter constraint is enforced on average, which seems to be unavoidable in mean-field treatments; as pointed out in Ref.\,\onlinecite{Watanabe2007}, this may restrict analyses on the character of the Fermi surface, as far as being hole-like or electron-like, large or small. 
However, our main purpose here is to gain insight into the stabilization of different magnetic modes, so that the tradeoff justifies imposing the constraint in its weaker form. 

As discussed in the Appendix, after substituting Eqs.\,\eqref{Si}-\eqref{triplet_hyb_mean_v} in the mean-field Hamiltonian, Eq.\,\eqref{hamil2},  imposing periodic boundary conditions, and performing a discrete Fourier transform, we obtain our working Hamiltonian, Eq.\,\eqref{FT_hamil4x4}. 
It is represented by a $4 \times 4$ matrix, which can be straightforwardly diagonalized, leading to the bands $ E^{n}_{\mathbf{k}},\ (n =~1,\ldots,4)$.

The Helmholtz free energy then becomes
\begin{equation}
F = -\frac{1}{\beta} \sum_{n, \mathbf{k}} \ln \big( 1 + e^{-\beta E^{n}_{\mathbf{k}}} \big) + const ,
\end{equation}
where $\beta=1/k_{B}T$; $k_{B}=1$ throughout this paper. 
The effective fields $V_{0}$, $V_{0}'$, $m^{0}_f$, $m^{0}_c$, $\mu$, $\epsilon_{f}$, and $ \mathbf{Q}$ are to be determined self-consistently by minimizing the Helmholtz free energy
\begin{align} \label{selfcon}
%line 1
\nonumber  &\bigg\langle \frac{\partial F}{\partial m_{f}^0} \bigg\rangle = \bigg\langle \frac{\partial F}{\partial m_{c}^0} \bigg\rangle = \bigg\langle \frac{\partial F}{\partial V_{0}} \bigg\rangle = \bigg\langle \frac{\partial F}{\partial V_{0}'} \bigg\rangle  \\ 
%line2
 &=\bigg\langle \frac{\partial F}{\partial \varepsilon_{f}} \bigg\rangle = \bigg\langle \frac{\partial F}{\partial \mu} \bigg\rangle = \bigg\langle \frac{\partial F}{\partial q_{x}} \bigg\rangle = \bigg\langle \frac{\partial F}{\partial q_{y}} \bigg\rangle = 0.
\end{align}
The resulting nonlinear coupled equations are solved numerically, using standard library routine packages, with the aid of the Hellmann-Feynman theorem.

At this point some comments are in order. 
First, we should mention that we have tried to include additional modulations to the field amplitudes, e.g., $ | \langle \mathbf{S}_{i} \rangle | \to m_f(\mathbf{R}_{i}) = m_{f}^{0} + m_{f}^{1} \cos\left( \mathbf{Q}\!\cdot\! \mathbf{R}_{i} \right)$, and similarly for $| \langle \mathbf{s}^{c}_{i} \rangle|$, $ | \langle V \rangle|$ and $ | \langle \mathbf{V} \rangle|$, but, as it turned out, the most stable solution always yields $m_f^1=m_c^1=V_1=V_1'=0$ for all ranges of $J/W$ and $n_{c}$ considered. 
Second, attempts to consider different $\mathbf{Q}$'s for any of $\langle \mathbf{S}_{i} \rangle$, $\langle \mathbf{s}^{c}_{i} \rangle$ and  $\mathbf{V}$ amount to a much harder minimization procedure, and have led either to unphysical results, such as spin amplitudes larger than 1/2, or to trivial mean-field solutions.

%%%%%%%%%%%%%%%%%%%%%%   RESULTS   %%%%%%%%%%%%%%%%%%%%%%%%%%%%%%%%%%%%%%%%

\section{\label{sec3} Ground State behaviour}

\begin{figure}[t]
\includegraphics[scale=0.22]{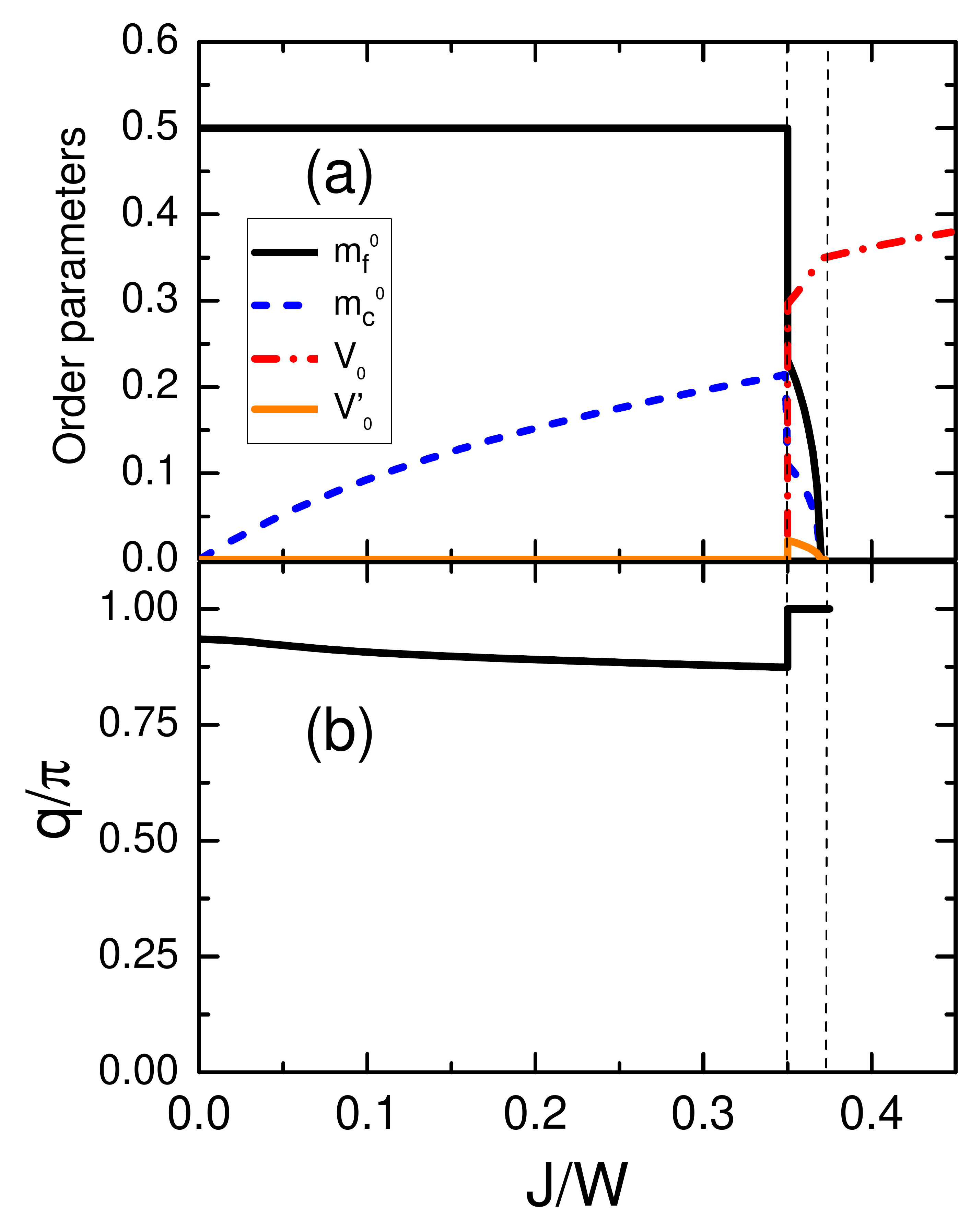}
\caption{(Colour online) (a) Order parameter amplitudes as functions of the Kondo exchange coupling $J$ (in units of the bandwidth, $W=8t$): local moment, $m_{f}^{0}$ (full black line), conduction electron magnetization, $m_{c}^{0}$ (dashed blue line), singlet hybridization, $V_{0}$ (dashed-dotted red line) and triplet hybridization, $V_{0}'$ (full orange line). 
(b) Magnetic wavevector component $q$ as a function of $J/W$; the spiral magnetic phase is described by the wavevector $\mathbf{Q}=(\pi,q)$. 
All data are for conduction electron density $n_c=0.90$, and zero temperature. 
}
\label{090}  %Fig 1
\end{figure}

\begin{figure}[t]
\includegraphics[scale=0.22]{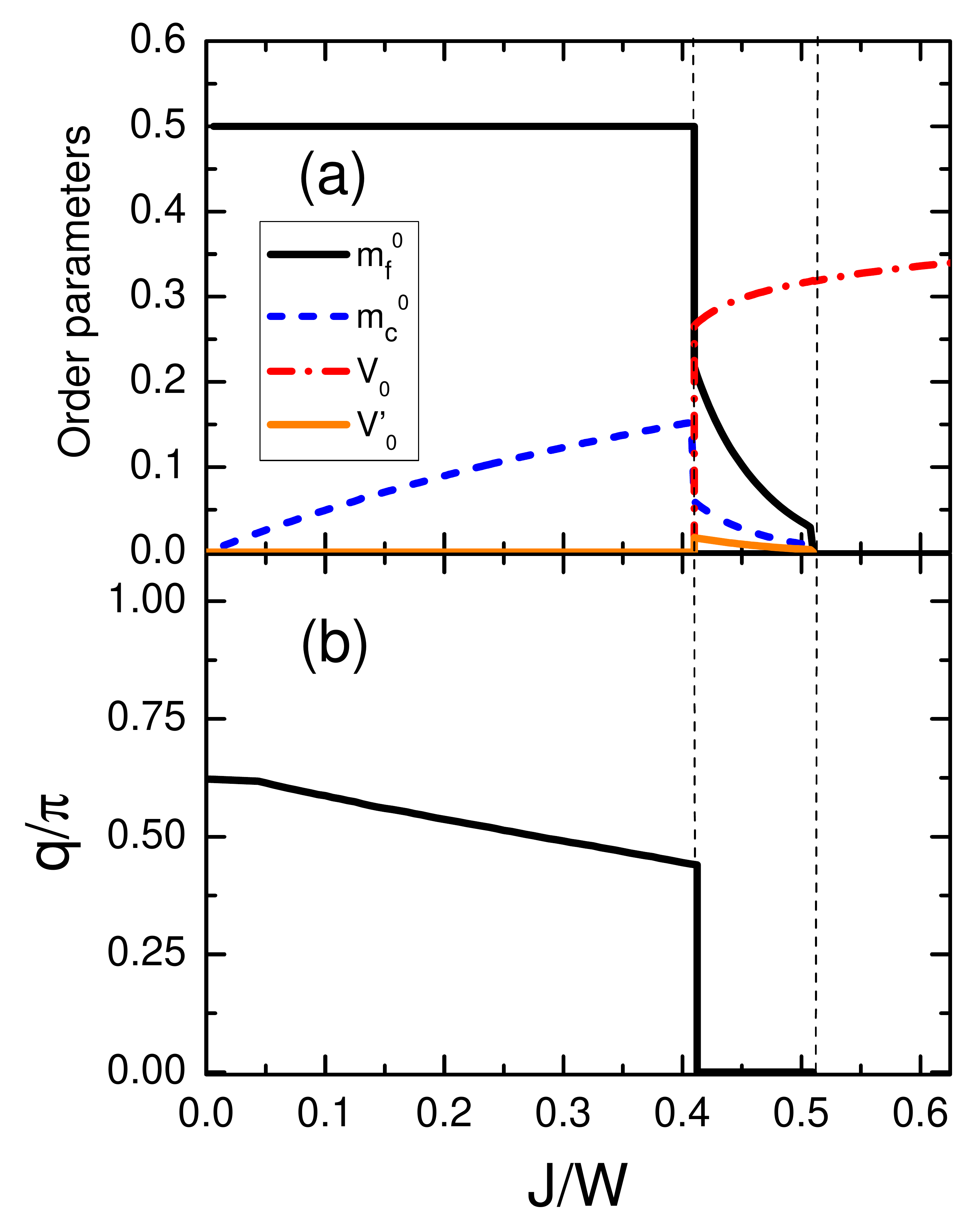}
\caption{(Colour online) Same as Fig.\,\ref{090}, but for $n_{c}=0.60$.}
\label{060} %Fig 2
\end{figure}

\begin{figure}[b]
\includegraphics[scale=0.22]{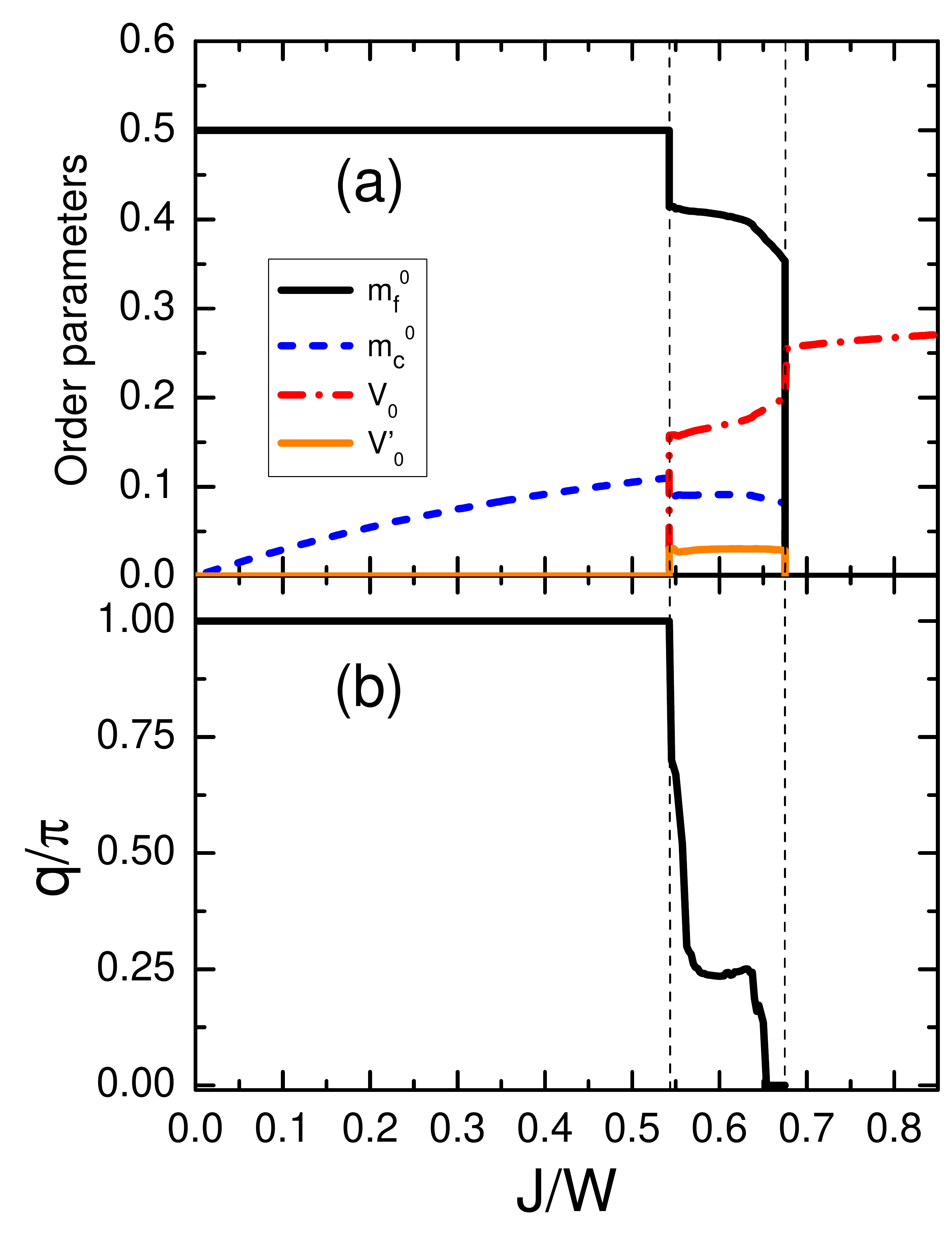}
\caption{(Colour online) Same as Fig.\,\ref{090}, but for $n_{c}=0.35$. 
}
\label{035} %Fig 3
\end{figure}

\begin{figure}[t]
\includegraphics[scale=0.22]{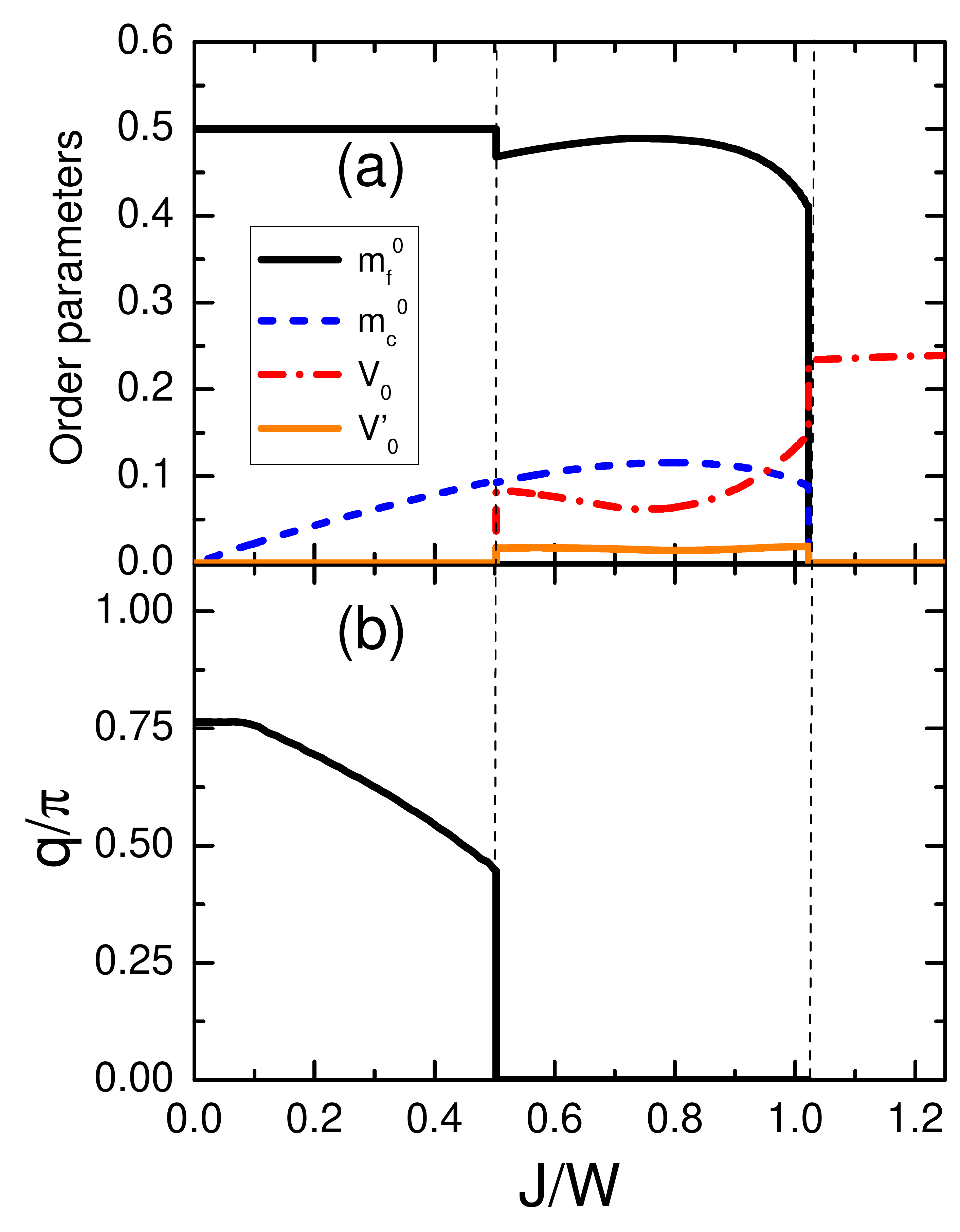}
\caption{(Color online) Same as Fig.\,\ref{090}, but for $n_{c}=0.25$, and now the magnetic wavevector is $\mathbf{Q}=(0,q)$.
}
\label{025} %Fig 4
\end{figure}

The set of nonlinear coupled equations, Eq.\,\eqref{selfcon}, is solved numerically for each pair of ($n_{c},J$), by fixing the temperature ($T=0$, for the time being) and the electronic density, and by letting  the exchange coupling to vary. 
Figures \ref{090} to \ref{025} show the behaviour of the order parameter amplitudes, Eqs.\,\eqref{Si}-\eqref{triplet_hyb_mean_v}, 
as functions of the exchange coupling $J$ (in units of the bandwidth, $W=8t$) for different doping levels (i.e., $n_{c}<1$). 
The figures also display the behaviour of the magnetic wavevector with $J/W$: as we will see, the magnetic modes and the coexistence with Kondo screening depend strongly on the electronic density. 

At half filling, the system is known to be an insulator for all $J/W$, but a quantum phase transition between an antiferromagnetic state and a spin singlet takes place at $(J/W)_c\simeq 0.4$.
Care must be taken when comparing this estimate with those of Refs.\,\onlinecite{Assaad99,Capponi01,Jurecka01}, since their \emph{working} Hamiltonians (i.e., after some decouplings or effective Hamiltonians are introduced) is somewhat different from ours, Eq.\,\eqref{FT_hamil4x4}. 
Nonetheless, a rough correspondence can be worked out from which the exchange coupling $J$ used in those works is one half of the one we use here; with this proviso, our estimate for $(J/W)_c$ is in good agreement with those of Refs.\,\onlinecite{Assaad99,Capponi01,Jurecka01}.

Figure \ref{090}(a) shows the results for $n_{c}=0.90$. 
In the weak-coupling regime, there is no hybridization (hence no Kondo screening); the local-moment amplitude is not affected by the exchange, while the amplitude of the conduction electron magnetization 
increases steadily with $J/W$.
The stable magnetically ordered phase corresponds to a spiral arrangement with wavevector $\mathbf{Q}=(\pi,q) $ [or, by symmetry, with $(q,\pi)$], and from Fig.\,\ref{090}(b), we see that $q$ first decreases slightly with $J/W$, hence drifting away from the N\'eel case; that is, the increase in the Kondo coupling by itself cannot drive the system into an antiferromagnetic state. 
However, when $J/W \approx 0.35$, hybridization abruptly sets in, causing a sudden decrease in both magnetic amplitudes; the magnetic mode also changes abruptly, stabilizing an AFM phase, with $\mathbf{Q}=(\pi,\pi)$, which coexists and competes with Kondo screening. 
Further increase in $J/W$ enhances the singlet hybridization which, in turn, steadily suppresses the magnetic amplitudes, both vanishing at $J/W \approx 0.375$.
The triplet hybridization amplitude $V_{0}'$ also vanishes at $J/W \approx 0.375$, tracking the suppression of the magnetization.
Beyond this point, there is only a paramagnetic phase with non-zero $V_{0}$, usually referred to as the Kondo phase. 
One can also see from Fig.\,\ref{090}(a) that the order parameters are discontinuous across the lower transition (into the coexistence region), and continuous at the second transition, into the screened-only (Kondo) phase.
The most stable ground state therefore corresponds to $V_{0}' \ll V_{0}$, which indicates that the competition with magnetism is almost entirely due to the singlet hybridization; accordingly, from now on hybridization effects will only be associated with $V_{0}$.

Moving on to $n_c=0.60$, we see from Fig.\,\ref{060}(a) that the order parameters behave in a way similar to the case with $n_c=0.90$, including the order of the transitions; in addition, a similar magnetic mode with $\mathbf{Q}=(\pi,q)$ is stabilized in this case. 
As shown in Fig.\,\ref{060}(b), $q$ also decreases with $J/W$ in the unscreened region, though with the important difference that in the coexistence region it is the $(\pi,0)$ mode which dominates. 
Later on, we will discuss the behaviour of $\mathbf{Q}$ as a function of $n_c$, for fixed $J/W$.  

\begin{figure}[t]
\includegraphics[scale=0.30]{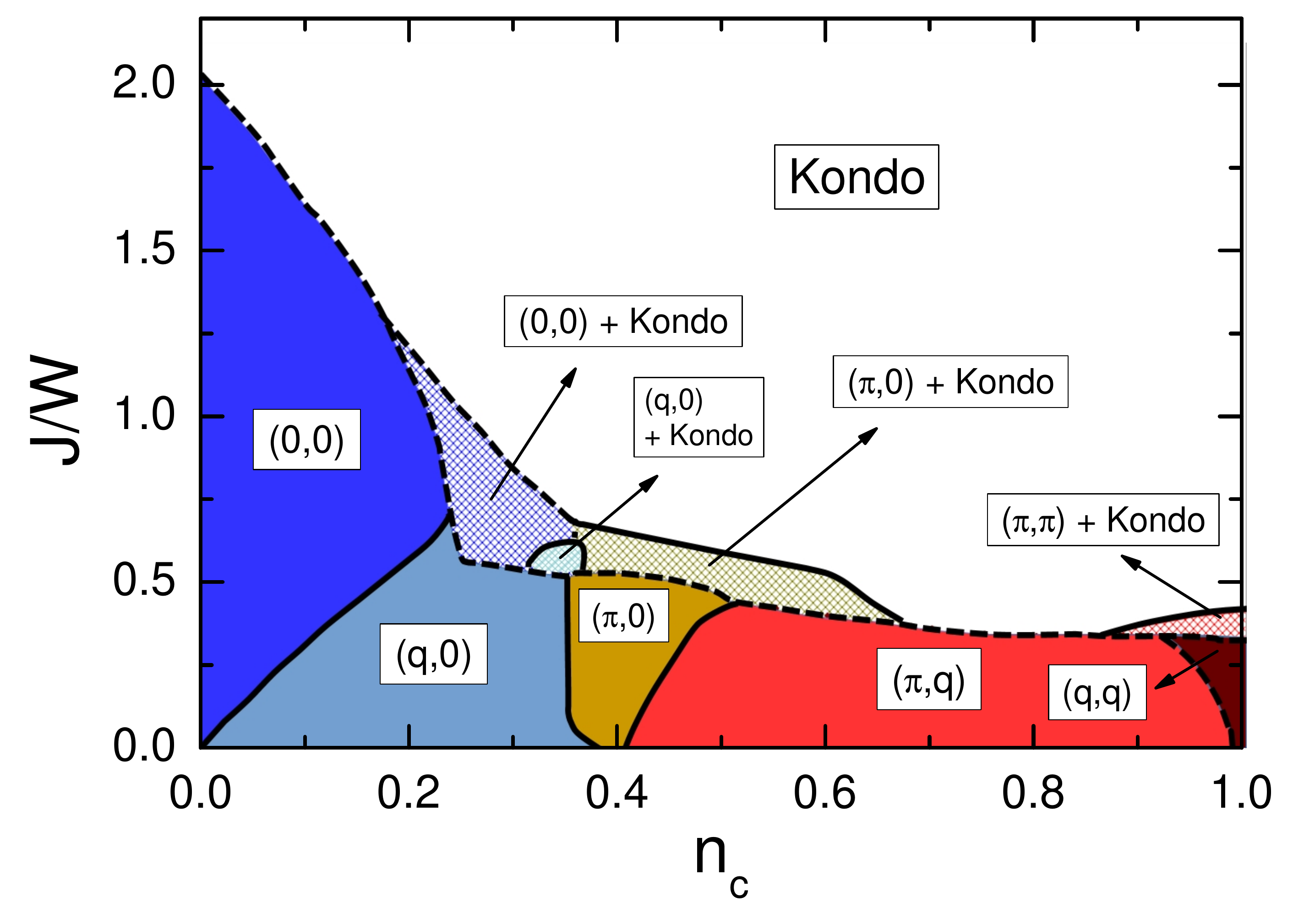}
\caption{(Color online) The ground state phase diagram, Kondo exchange \emph{versus} electronic density.  
The magnetic phases are designated by their magnetic wavevectors, $\mathbf{Q}=(q_x,q_y)$, and `Kondo' denotes a phase in which screening, as measured by the hybridization, is present (see text). Solid and dashed lines respectively represent continuous and discontinuous phase transitions. 
}
\label{ph_diag} %Fig 5
\end{figure}

Figure \ref{035} shows the corresponding analysis for $n_{c} = 0.35$. 
We see that the stable magnetic phase now has a wavevector $\mathbf{Q} = (\pi,0)$ in the unscreened region. 
In the coexistence region, the wavevector first stabilizes in a mode $\mathbf{Q} = (q,0)$, before becoming FM, $\mathbf{Q} = (0,0)$, as showed in Figure \ref{035}\,(b); see also Fig.\,\ref{ph_diag}. 
To the best of our knowledge, this is the first time that the coexistence of a spiral incommensurate magnetic mode with the Kondo phase is predicted within a static mean-field analysis; more on this coexistence later.

Further decrease in the electronic density, e.g., for $n_{c}=0.25$, leads to a spiral magnetic phase with $\mathbf{Q}=(q,0)$; see Fig.\,\ref{025}. 
While in the unscreened phase one finds a monotonically decreasing $q(J/W)$, in the coexistence region a uniform FM phase [i.e., one with $\mathbf{Q}=(0,0)$] is stabilized. 
Interestingly, while the transition into coexistence (which occurs at $J/W \approx 0.51$) is still of first order, here we see that, unlike what we have discussed so far, the transition to pure Kondo behaviour, occurring at $J/W \approx 1.02$ is also discontinuous. 

Similar analyses were performed for other values of $n_{c}$, the results of which are summarized in the phase diagram of Fig.\,\ref{ph_diag}. 
In line with previous mean-field approaches \cite{Lacroix1979,Fazekas1991}, we see that at low densities a saturated ferromagnetic phase is stable, while at half filling it is an antiferromagnetic phase which is the stable one.
On the other hand, we have established that the evolution of magnetic modes with the electronic density is much smoother than hitherto assumed; the diagram of Fig.\,\ref{ph_diag} also shows that magnetically ordered phases can still withstand some screening, though no trace of magnetism is found deep in the Kondo phase, as expected. 

\begin{figure}[b]
\includegraphics[scale=0.3]{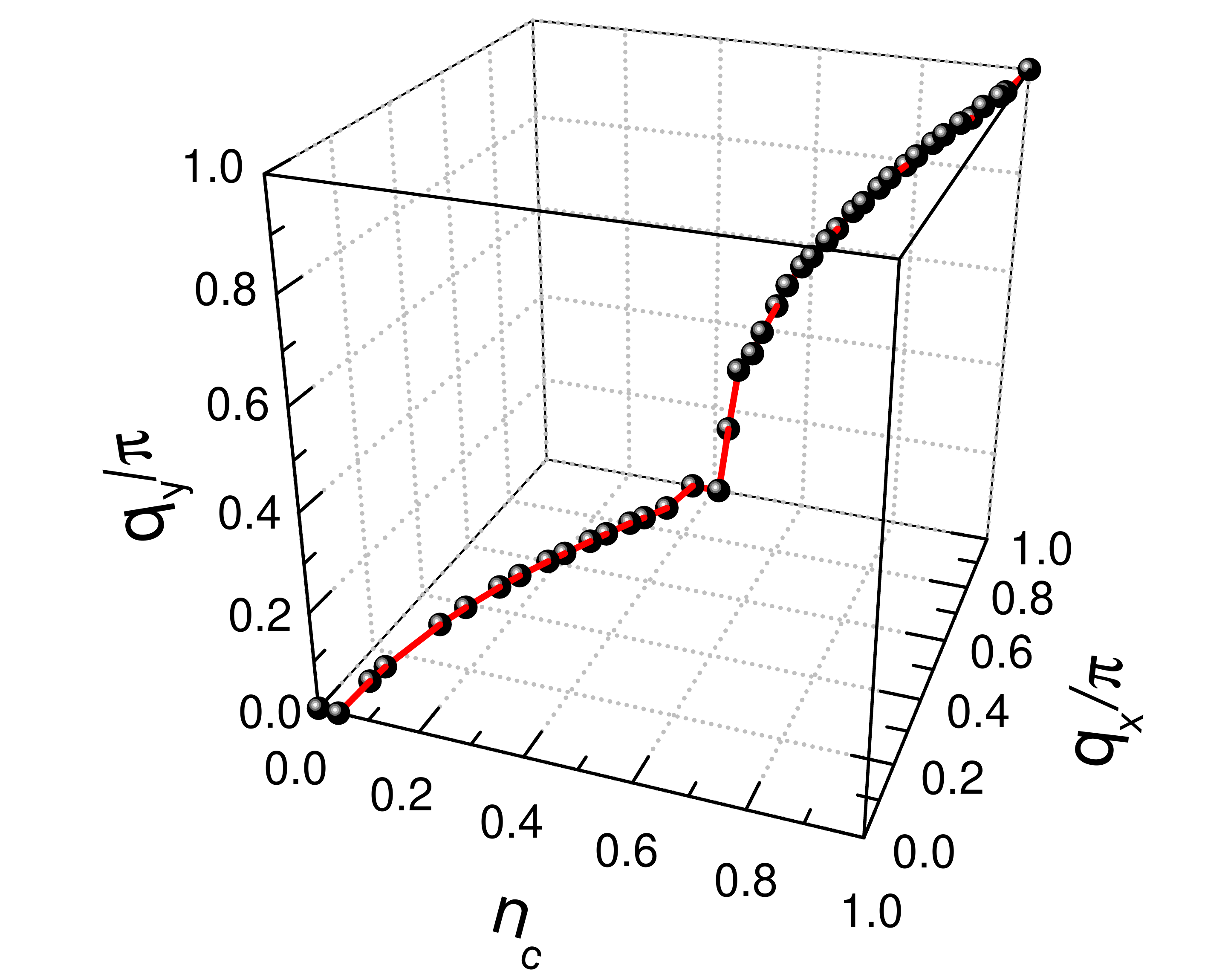}
\caption{(Color online) 
The magnetic wavevector $\mathbf{Q}=(q_x,q_y)$ as a function of electron density, for fixed $J/W=0.125$.
}
\label{q-dependence} % Fig 6
\end{figure}

We now discuss these aspects in turn, starting with the magnetic ordering in the absence of Kondo screening.
As the electronic density increases from zero, a spiral modulation develops in one of the lattice directions, say, the $x$-direction, while the same modulation is repeated along the $y$-direction, for which $q_y=0$. 
Figure \ref{q-dependence} illustrates the evolution of the modulation vector with the density, for fixed $J/W$: the modulation along $x$ becomes staggered ($q_x=\pi$), and so remains, even as the density increases slightly up to $n_c\sim 0.4$. 
As $n_c$ continues to increase, the modulation along $y$ starts changing until it also reaches $q_y=\pi$ close to half filling.

The presence of spiral magnetic phases has been found for the one-dimensional KLM \cite{Garcia2004}, in addition to ferromagnetic `island' phases; while our results reveal that spiral phases also occur for the square lattice, no ferromagnetic island states led to minimal free energies in the present case. 
The continuous change in one of the components of the magnetic wavevector can be attributed to a distribution of conduction electrons preferentially along one of the lattice directions; similar effects  have been observed in the DMRG study of the 2-leg Kondo ladder \cite{Xavier2004}; that is, the Kondo lattice seems to develop a stripe structure.    
The correspondence with the 2-leg Kondo ladder goes even farther: fits to the linear portions of $q(n_c)$ in Fig.\,\ref{q-dependence} yield $q\approx n_c$ and $q\approx 2n_c$, respectively near half-filling, and near $n_c\sim 0.35$. 
This should be compared with $\mathbf{Q}=(n_{c},1)\pi$ and $\mathbf{Q}=(2n_{c},0)\pi$, for $n_{c} \gtrsim 0.5$ and $n_{c} \lesssim 0.5$ respectively, for the ladder \cite{Xavier2004}.
On the other hand, there is a noticeable difference for a range of densities near $n_c \approx 0.4$, where, for the square lattice, the commensurate phase $(\pi,0)$ stabilizes; see Fig.\,\ref{ph_diag}.

\begin{figure}[t]
\includegraphics[scale=0.30]{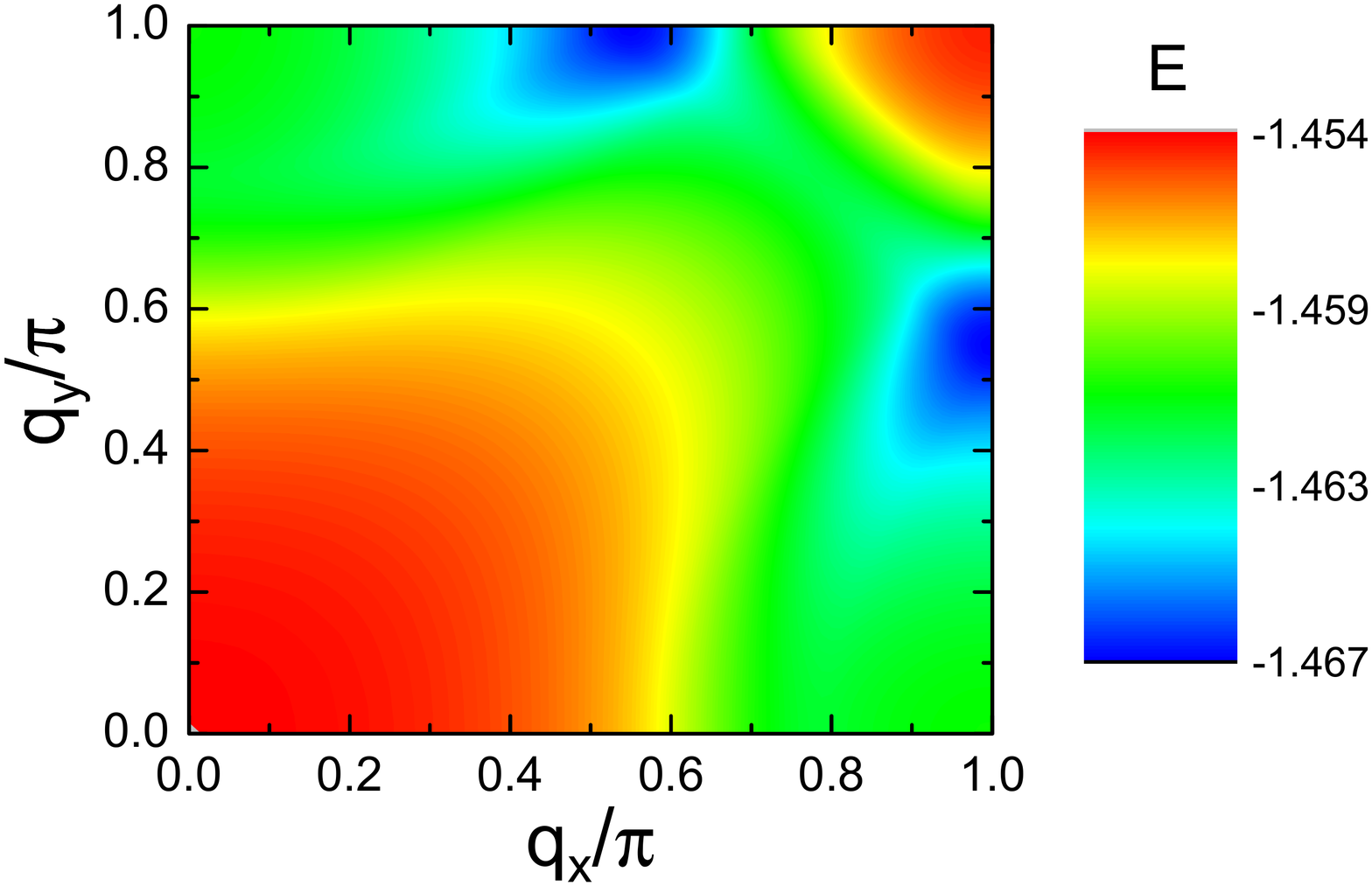}
\vskip -0.5cm
\caption{(Color online) Internal energy contour map as a function of the magnetic wavevector, for $n_{c}=0.60$ and $J/W=0.175$. 
}
\label{energy} % Fig 7
\end{figure}

\begin{figure}[h]
\includegraphics[scale=0.30]{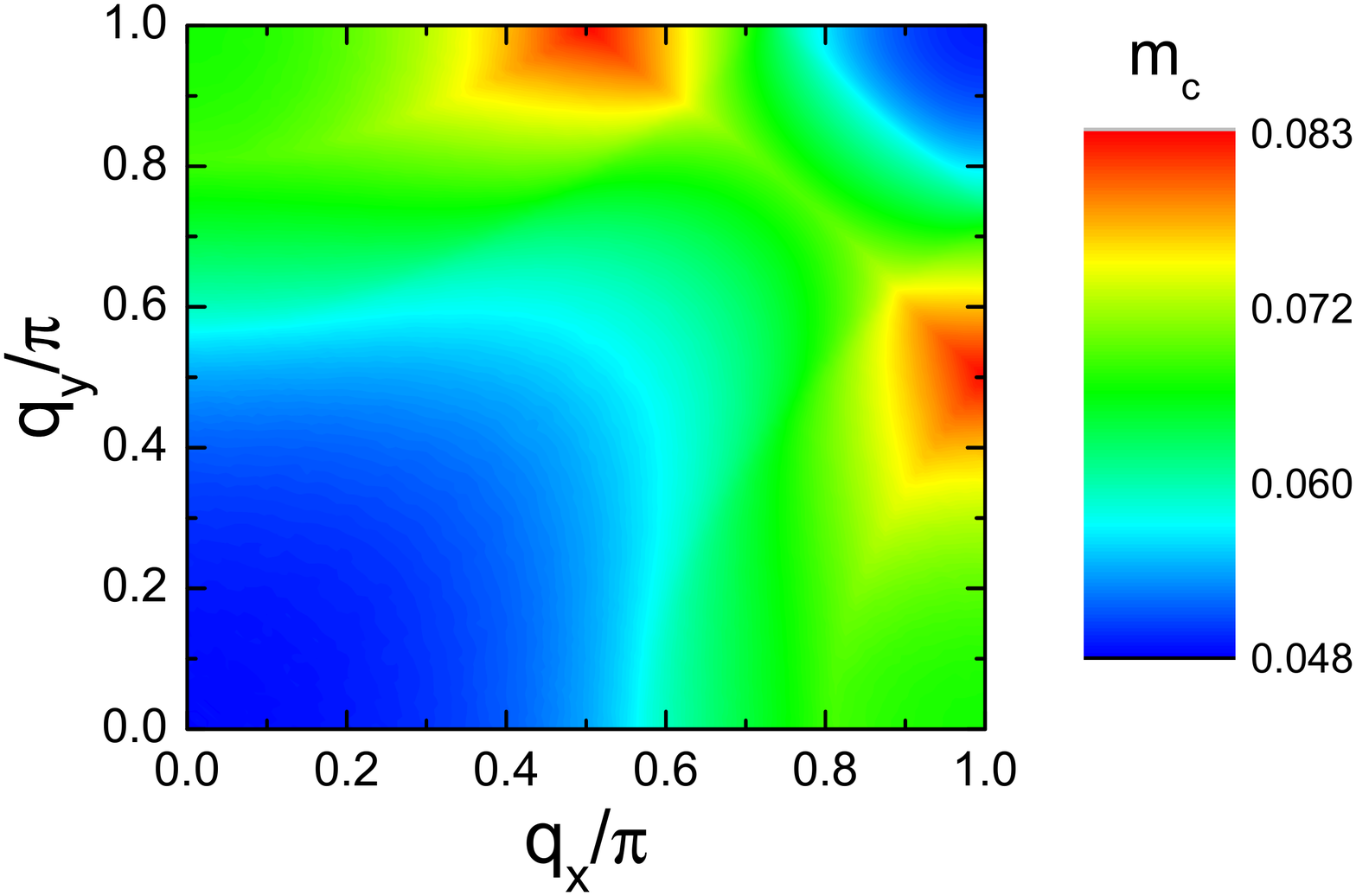}
\vskip -0.5cm
\caption{(Color online) Polarization contour map as a function of magnetic wavevector, for the same parameters as in Fig.\,\protect{\ref{energy}}. 
}
\label{polar} % Fig 8
\end{figure}

Let us now compare these predictions with the experimental data for the borocarbide alloy Tb(Co$_{x}$Ni$_{1-x}$)$_{2}$B$_{2}$C [Refs.\,\onlinecite{ElMassalami2012,ElMassalami2013}].  
The planar magnetic arrangement in TbNi$_{2}$B$_{2}$C is a spin-density wave, with a wavevector close to $(\pi,0)$ [Ref.\onlinecite{Lynn97}], so that we can represent this compound by the $n_c\approx 0.4$ point in Fig.\,\ref{q-dependence}. 
Assuming the primary effect of the gradual substitution of Ni by Co is a decrease in the number of conduction electrons (hence of $n_c$), Fig.\,\ref{q-dependence} correctly predicts that the alloy evolves towards a saturated ferromagnet in the opposite limit of 100\% Co; the comparison cannot be made for intermediate dilutions, since partial replacement mainly affects the modulation along the $c$-axis.
In the corresponding case of the Ho alloys, the planar arrangement is ferromagnetic for all Co concentrations, $x$, while the magnetic modulation along the $c$-axis is strongly dependent on $x$.\cite{ElMassalami2014} 
However, since Ho(Co$_{x}$Ni$_{1-x}$)$_{2}$B$_{2}$C is superconducting below $x=0.03$, one expects electron-electron interactions to play an important role in the ensuing magnetic arrangement, even when the system is not superconducting.
While these effects are certainly absent in the simple model considered here, the capture of the trend observed in the Tb alloys may be taken as an indication that the KLM is a viable starting point to describe the magnetism in this class of materials.

Further insight can be gained by discussing the relative stability between AFM and FM phases. 
Figure \ref{energy} shows a contour map of the internal energy $E$ as a function of magnetic wavevector ($q_{x}, q_{y}$), for $J/W=0.175$ and $n_{c}=0.60$. 
The map is obtained by minimizing the energy, Eq.\,\eqref{selfcon}, with respect to all variables, but $q_{x}$ and $q_{y}$.
From Fig.\,\ref{energy} we see that when $ \mathbf{Q} = (\pi, \pi)$ and (0,0), the internal energy reaches its largest values, showing that for this choice of ($n_{c},J/W$) the most stable magnetic arrangement in the ground state is neither AFM nor FM. 
The minimum of the internal energy actually occurs for ($q,\pi$) [and, by symmetry for ($\pi,q$), as well], with $ q/\pi \approx 0.55 $, thus providing us with an explicit example showing that many different magnetic arrangements may be closely separated in energy. 
Figure \ref{polar} shows the corresponding contour map for the polarization, $m_{c}^{0}$. 
We first note that the polarization is in opposite phase in relation to the internal energy. 
For this choice of ($n_{c},J/W$), the hybridization is zero (see Fig.\,\ref{060}), and the system is dominated by the RKKY interaction, so that the polarization is the sole driving force to magnetism. 

\begin{figure}[t]
\includegraphics[scale=0.23]{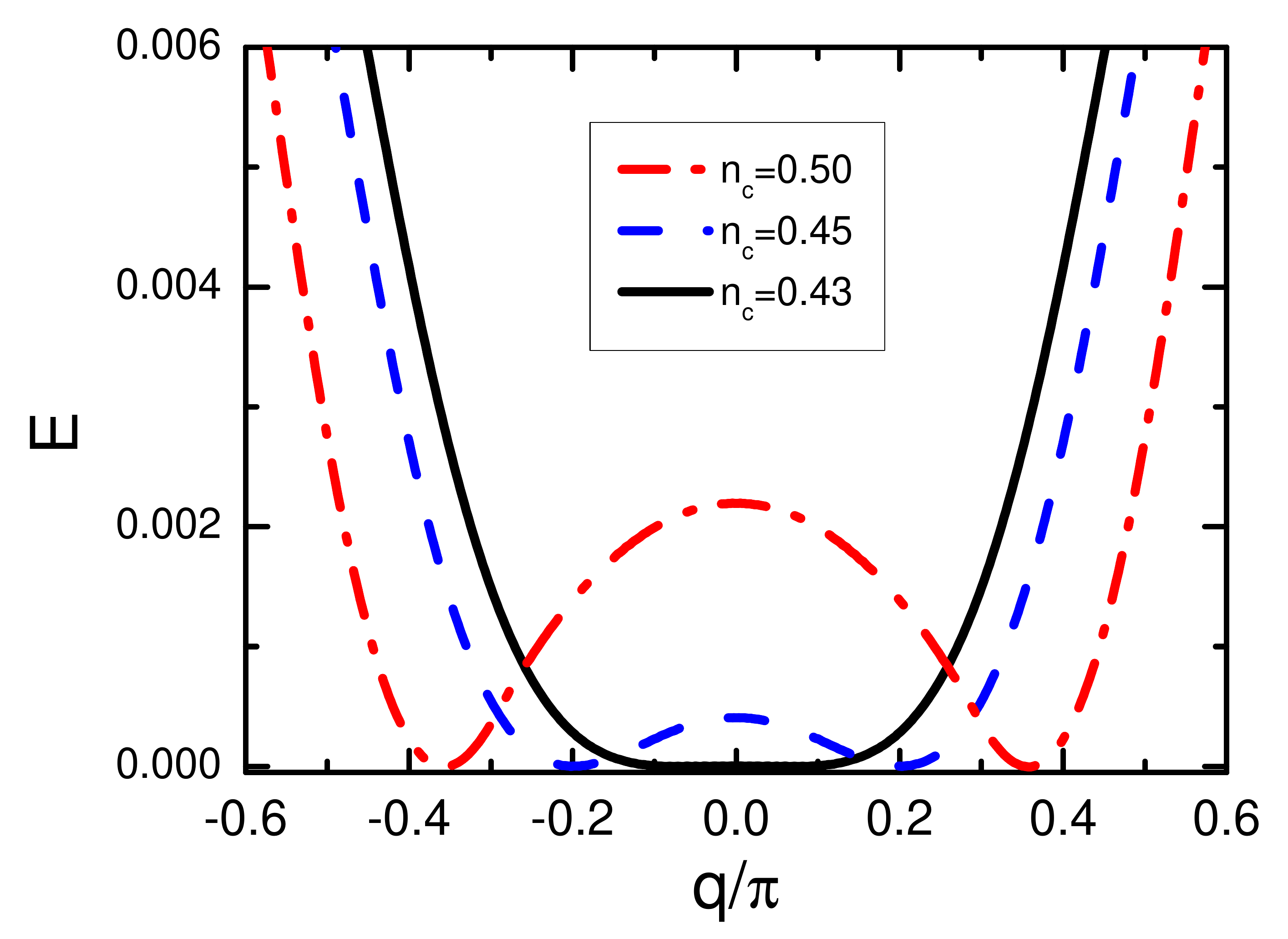}
\caption{Internal energy as a function of magnetic wavevector as the zero-temperature transition from $(\pi,q)$ to $(\pi,0)$ is approached by varying the electronic density, $n_c$, for fixed $J/W=0.125$. The curves were shifted for easier comparison.}
\label{2nd_ord} % Fig 9
\end{figure}

As far as the order of the density-driven transitions is concerned, we should add a few comments. 
First, we note that the transition is necessarily discontinuous if $\mathbf{Q}$ changes abruptly at the boundary; as shown in Fig.\,\ref{q-dependence}, $\mathbf{Q}$ does not suffer any discontinuity in the whole range of $n_c$. 
Further evidence comes from Fig.\,\ref{2nd_ord}, which shows the internal energy as a function of $q$, as the transition from $(\pi, q)$ to $(\pi,0)$ is approached, for fixed $J/W$ and varying $n_{c}$: we see that the two global minima go continuously to zero as the critical point is approached. 
Analogous behaviour occurs for all other transitions at fixed $J/W$. 

We now discuss the coexistence of Kondo screening and magnetically ordered phases. 
First, we note that this coexistence is restricted to moderate degrees of screening, as indicated in Figs.\,\ref{090} to \ref{025}, and summarized in Fig.\,\ref{ph_diag}.
In line with other mean-field analyses \cite{Zheng1996, Zhang2000, Zhang2010,Asadzadeh2013}, our results show that close to half filling, i.e., for $n_{c} \gtrsim 0.86$, Kondo screening coexists with an AFM mode, while for low electronic densities ($ 0.15 \lesssim n_{c} \lesssim$ 0.31) this coexistence occurs with the FM configuration. 
However, our approach allows us to go further, and establish that for intermediate electronic densities,  0.31 $\lesssim n_{c} \lesssim 0.68 $, coexistence is possible with phases other than FM and AFM; see Figs.\,\ref{060}, \ref{035}, and \ref{ph_diag}.
A commensurate magnetic phase, with wavevector $\mathbf{Q} = (\pi, 0)$, stabilizes into the coexistence region for the range $ 0.36 \lesssim n_{c} \lesssim 0.68 $. More interestingly, for the range $ 0.31 \lesssim n_{c} \lesssim 0.36 $ such coexistence is with a spiral incommensurate magnetic phase.

\begin{figure}[t]
\includegraphics[scale=0.20]{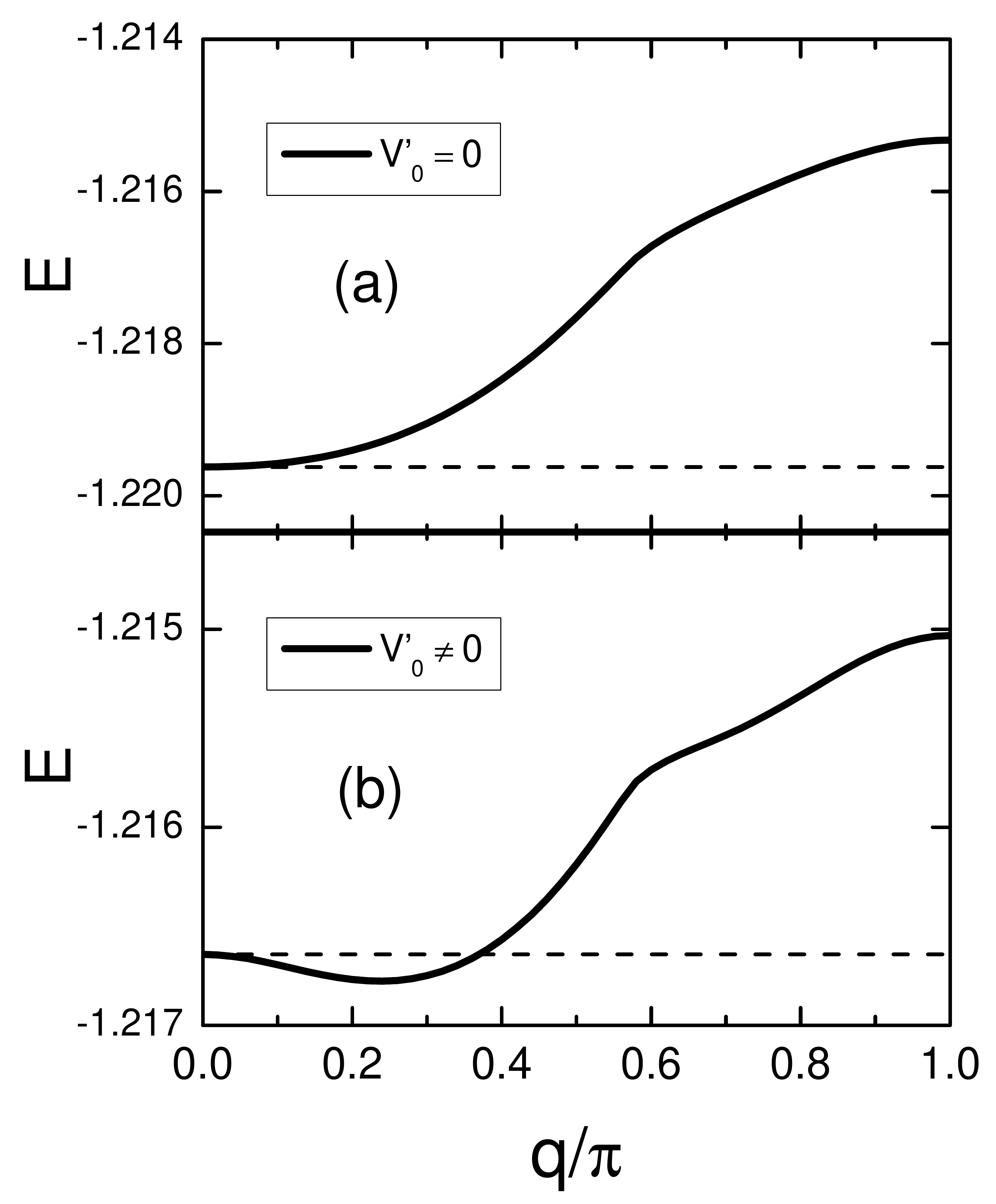}
\caption{Internal energy as a function of magnetic wavevector $\mathbf{Q}=(q, 0)$ at $n_{c}=0.35$ and for fixed $J/W=0.59$, in the cases where (a) the triplet hybridisation term ($V_{0}'$) is forced to be zero, and (b) when it is allowed to be non-zero, in the coexistence region. The thin black dashed line is the ferromagnetic internal energy for both cases.}
\label{Qlivre_035} % Fig 10
\end{figure}

Let us now focus on the coexistence between Kondo and spiral phases. 
Figure \ref{Qlivre_035} (a) compares the internal energy as a function of $q/\pi$ (for fixed $J/W$) when the triplet hybridisation amplitude ($V_{0}'$) is constrained to be zero, and (b) when it is allowed to be non-zero. 
In the former case, the energy is minimum at the ferromagnetic mode $\mathbf{Q}=(0, 0)$, while when this constraint is relaxed a mode with $q/\pi \approx 0.25$ becomes the most stable one. 
It should also be stressed that, by contrast, the final minimisation outcome for the remaining auxiliary fields, $m_{f}^{0}$, $m_{c}^{0}$ and $V_{0}$, is hardly affected by whether $V_{0}'$ is zero or non-zero.  
We conclude that the appearance of modes with $q\neq 0,\pi$ in the region of coexistence between Kondo screening and magnetic order is directly related to a modulation of the hybridisation with the magnetic wavevector $\mathbf{Q}$. 
Evidently, as $J/W$ varies, the value of $q$ which minimises the energy also varies; see Fig.\,\ref{035}\,(b).
Another subtle aspect is that the resulting amplitude of the modulated hybridisation is weak (typically $V_{0}'/V_{0} \simeq 0.1$), so that coexistence involving Kondo screening and modes with either $q\neq 0$ or $\neq\pi$ only occur in small portions of the diagram, the precise location of which would demand a much more elaborate analysis; suffices to say, for our purposes here, that coexistence with $q\neq 0,\pi$ is indeed possible.    

\begin{figure}[b]
\includegraphics[scale=0.23]{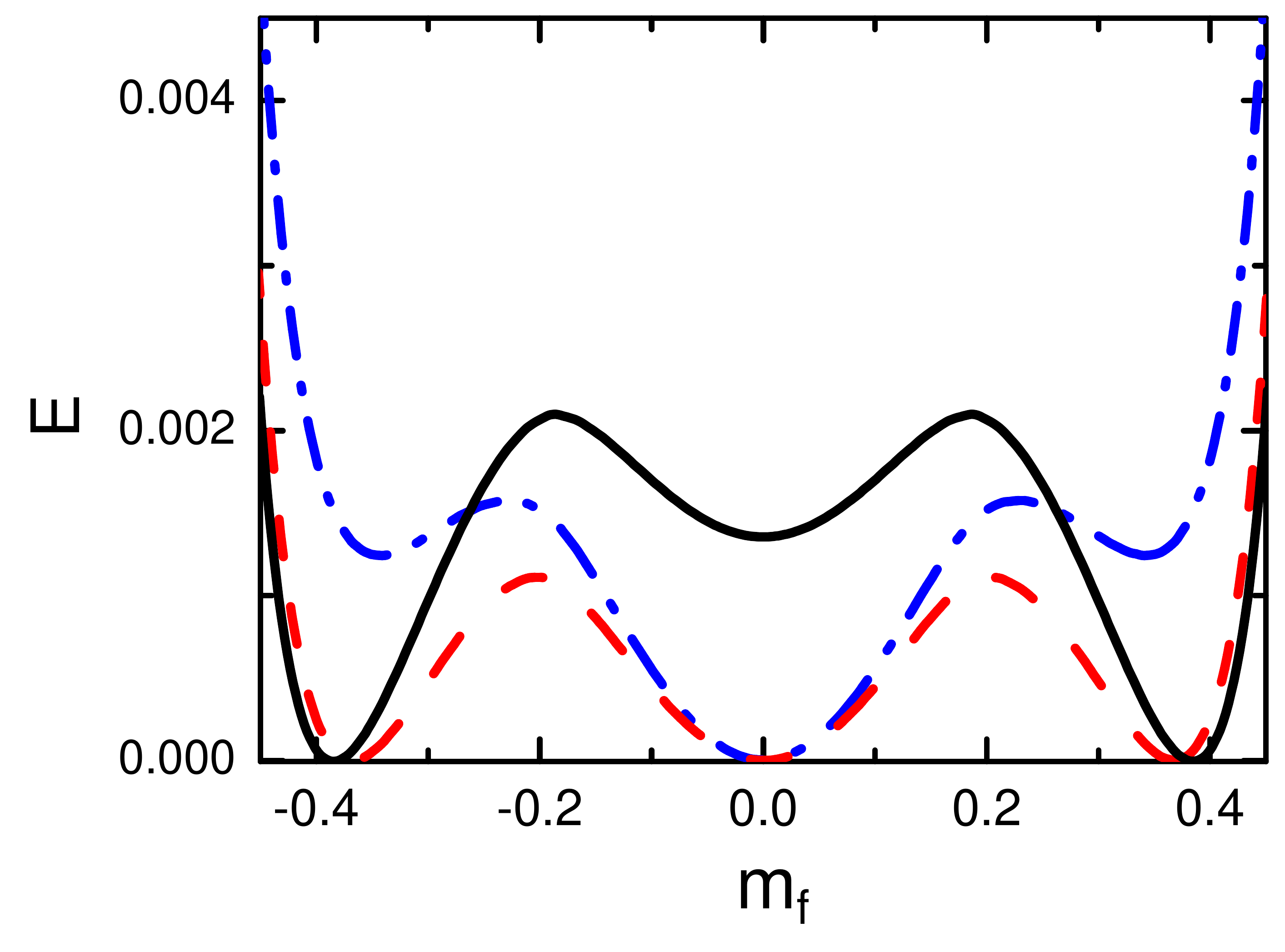}
\caption{Internal energy as a function of magnetization at $n_{c}=0.30$. The internal energy behaviour before ($J/W=0.85$), on ($J/W\approx0.866$) and after ($J/W=0.881$) the transition point to pure Kondo are represented by black (solid), red (dashed) and blue (dash-dotted) lines, respectively. The curves were shifted for easier comparison. }
\label{1st_ord} % Fig 11
\end{figure}

Finally, we examine the transition to the Kondo phase, which marks the disappearance of magnetism.
As shown in Figs.\,\ref{090}, \ref{060}, and \ref{ph_diag} the transitions to the pure Kondo regime from both the AFM+Kondo phase and from the $(0,\pi)$+Kondo phase, are continuous. 
By contrast, the transition from FM+Kondo is discontinuous, as it can be seen from Fig.\,\ref{1st_ord}, 
in which we fix the electronic density as $n_{c}=0.30$, and plot the internal energy as a function of the local moment amplitude $m_{f}^{0}$, for values of $J/W$ near the transition point, $(J/W)_c\approx 0.866$.
Within the coexistence region, $J/W = 0.85 < (J/W)_c$, the internal energy displays global minima at $m_{f}^{0} \approx \pm 0.4$, signalling a FM state, together with a local minimum at $m_{f}^{0}=0$.
At the transition point, this local minimum becomes degenerate with those for which $m_{f}^{0} \simeq \pm 0.4$, and for $ J/W = 0.881 > (J/W)_c$, the minimum at $m_{f}^{0}=0$ becomes the most stable one: the transition is therefore discontinuous.
Similar discontinuous behaviour of the internal energy is found for the direct transitions (i.e., without going through coexistence regions) to the Kondo phase from the $\mathbf{Q}=(0,0)$ (for $n_c\lesssim 0.15$) and $\mathbf{Q}=(\pi,q)$ ($0.5 \lesssim n_c \lesssim 0.85$) phases; see Fig.\,\ref{ph_diag}.
One should note that in Ref.\,\onlinecite{Zhang2010} the transition FM+Kondo to Kondo was found to be continuous; possible sources for this discrepancy may lie in either the constant density of states used in that work, or to the fact that the mean-field implementation differs from ours, especially for the FM solution (see the Appendix).
\begin{figure}[t]
\includegraphics[scale=0.30]{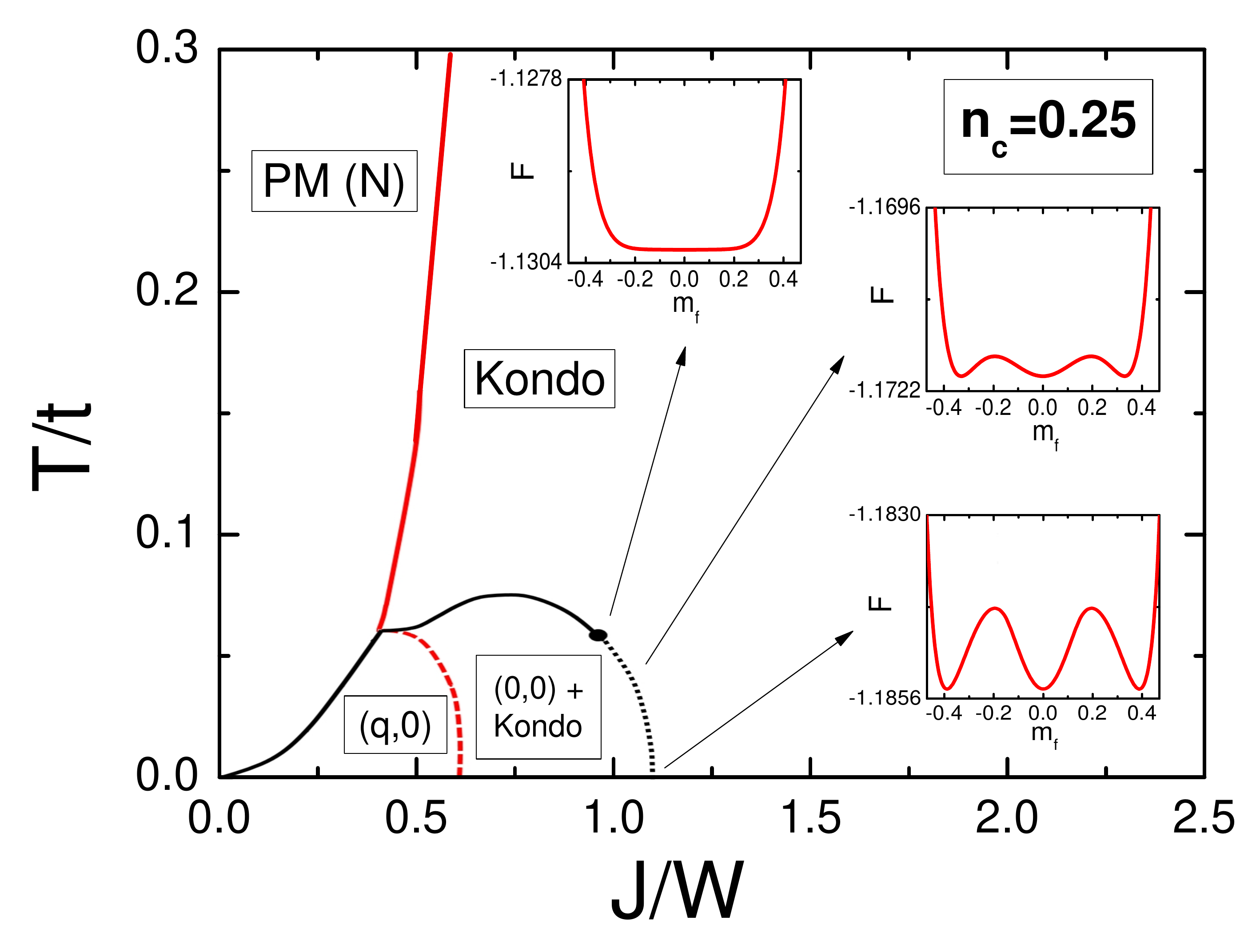}
\caption{(Color online) `Doniach diagram' for $n_{c}=0.25$. Solid lines represent continuous transitions, while broken (dashed and dotted) lines denote discontinuous transitions; red lines mark the onset of hybridization, while black lines mark the disappearance of magnetic order. 
PM(N) stands for normal (i.e., no hybridization) paramagnetic phase.
The dot in the boundary between FM+Kondo and Kondo phases marks the tricritical point; the insets show the behaviour of the Helmholtz free energy as a function of the average local moment, along the first-order boundary.}
\label{fig:Doniach025} % Fig 12
\end{figure}

%%%%%%%%%%%%%%%%%%%%%%   Finite temperature   %%%%%%%%%%%%%%%%%%%%%%%%%%%%%%%%%%%%%%%%

\section{\label{sec4} Finite Temperatures}

In analysing the behaviour at finite temperatures we fix $n_c$, while the temperature and the exchange coupling are allowed to vary.
Similarly to what we did for $T=0$, we examine the temperature dependence of the order parameters to determine the phase boundaries; we also determine the temperature dependence of the magnetic wavevector.

\begin{figure}[t]
\includegraphics[scale=0.3]{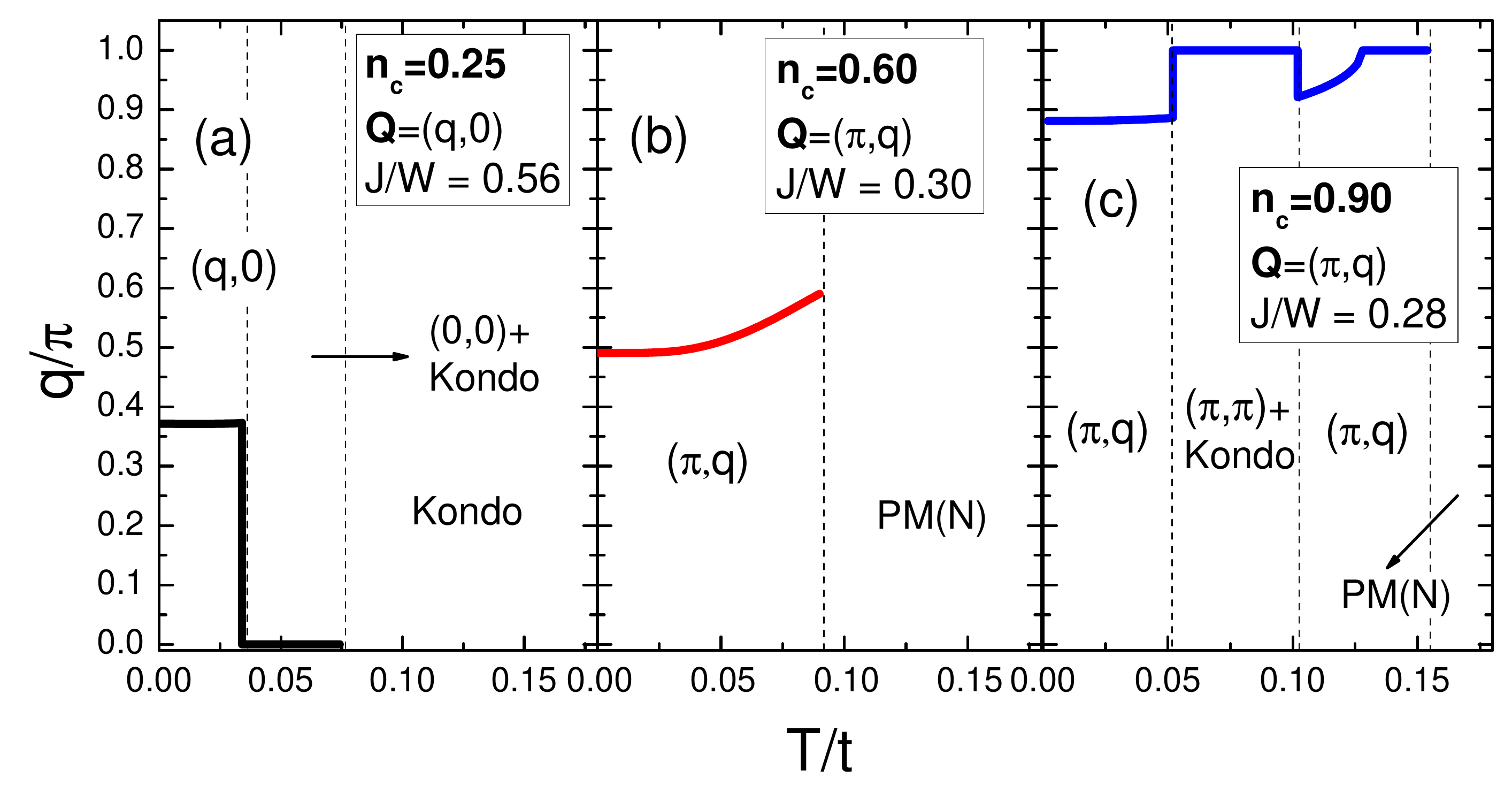}
\caption{(Color online) 
Temperature dependence of the variable wavevector component $q$, for three distinct combinations of $(n_c,J/W)$. The magnetic wavevector in each panel is given by (a) $\mathbf{Q}=(q,0)$, (b) $\mathbf{Q}=(\pi,q)$, and (c) $\mathbf{Q}=(\pi,q)$.}
\label{fig:qvsT} % Fig 13
\end{figure}

Figure \ref{fig:Doniach025} summarizes our findings for $n_c=0.25$ in the form of a `Doniach diagram',
in which several details on the nature of the magnetically ordered phases can now be unveiled. 
First, we note that for this electronic density, the unscreened magnetic phase is actually a spiral phase. 
More interestingly, the magnetic mode $\mathbf{Q}=(q,0)$ is such that $q$ displays a temperature dependence, as shown in Fig.\,\ref{fig:qvsT}(a), for a fixed $J/W= 0.56$. 
In this case, the magnetic mode is hardly dependent on the temperature in the unscreened region, but an abrupt change occurs as soon as the temperature drives the system into the coexistence region, where, in this case, ferromagnetism sets in over a temperature interval; for a slightly smaller value, say $J/W=0.5$, there is a noticeable temperature dependence of $q$ with $T$ (not shown), but the range of temperatures in which the FM phase exists is quite smaller than the one shown.  
The order of the transitions along the border between FM and PM Kondo phases changes from continuous (at higher temperatures) to discontinuous (lower temperatures); the insets show the evolution of the free energy along the first-order boundary, and at the tricritical point. 
For completeness, one should mention that a similar phase diagram was obtained in Ref.\,\onlinecite{Liu13} for a single electronic density, $n_c = 0.2$; however, since no spiral phases were considered there, the evolution of the wavevector $\mathbf{Q}$ with the temperature could not be established.

\begin{figure}[t]
\includegraphics[scale=0.30]{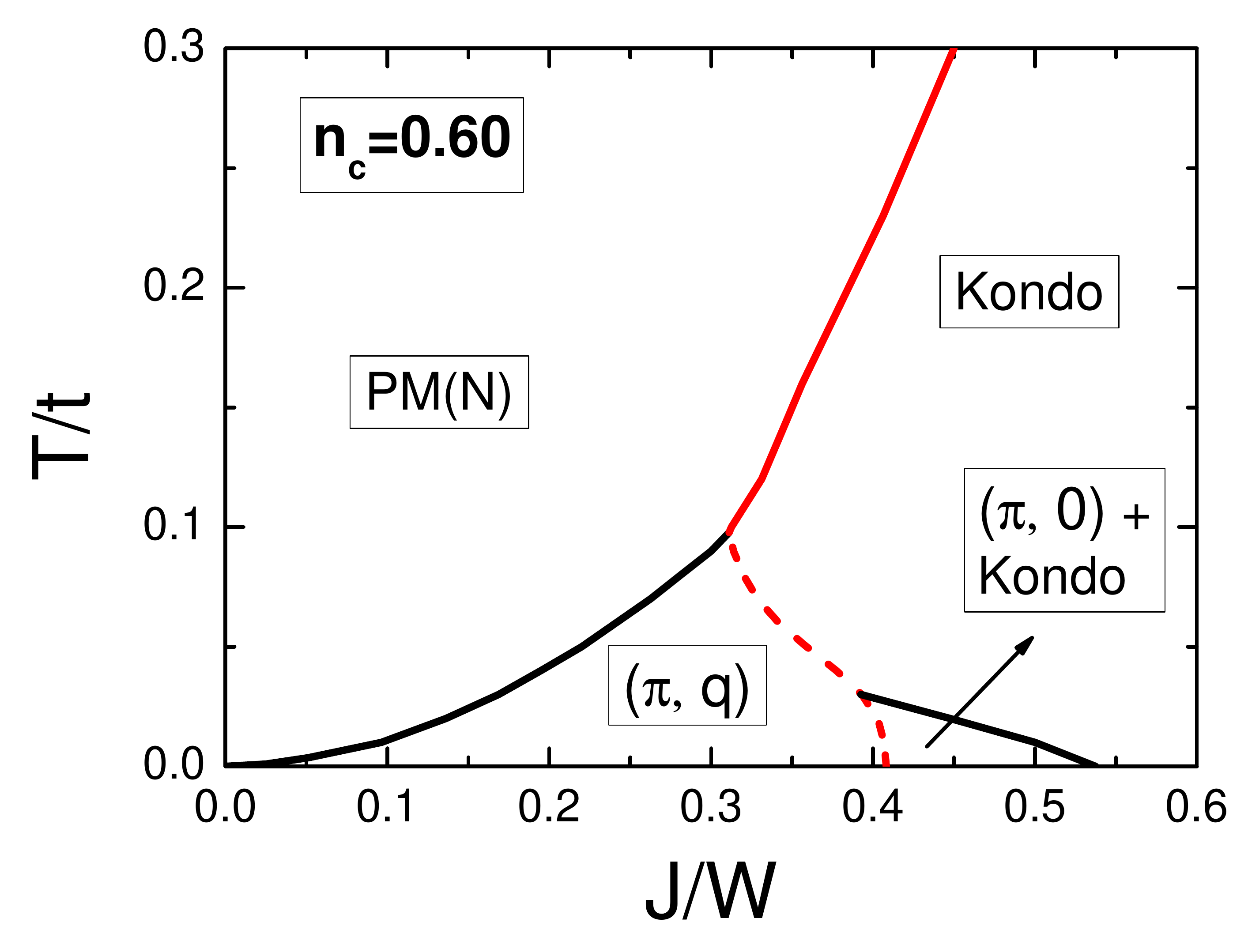}
\caption{(Color online) Same as Fig.\,\ref{fig:Doniach025}, but for $n_c=0.6$.}
\label{fig:Doniach06} % Fig 14
\end{figure}

The features brought about by the spiral phases also manifest themselves at other densities. 
Figure \ref{fig:Doniach06} shows data for $n_c=0.6$, in which case the spiral phases involved at $T=0$ are those for $\mathbf{Q}=(\pi,q)$, with $q$ decreasing as $J/W$ increases (see Fig.\,\ref{060}).
One notes that the magnetic boundary between $(\pi,0)$+Kondo and the Kondo phase is completely detached from the boundary between $(\pi,q)$ and PM(N); therefore, one can go from $(\pi,q)$ to the Kondo phase without an intervening coexistence region, simply by raising the temperature.
As Figure \ref{fig:qvsT}(b) shows, for $J/W=0.3$ the effect of temperature is to increase $q$, moving towards antiferromagnetism. 
On the other hand, the coexistence region which appears for $0.4 \lesssim J/W\lesssim 0.54$ involves the `striped' phase $\mathbf{Q}=(\pi,0)$.   

\begin{figure}[b]
\includegraphics[scale=0.30]{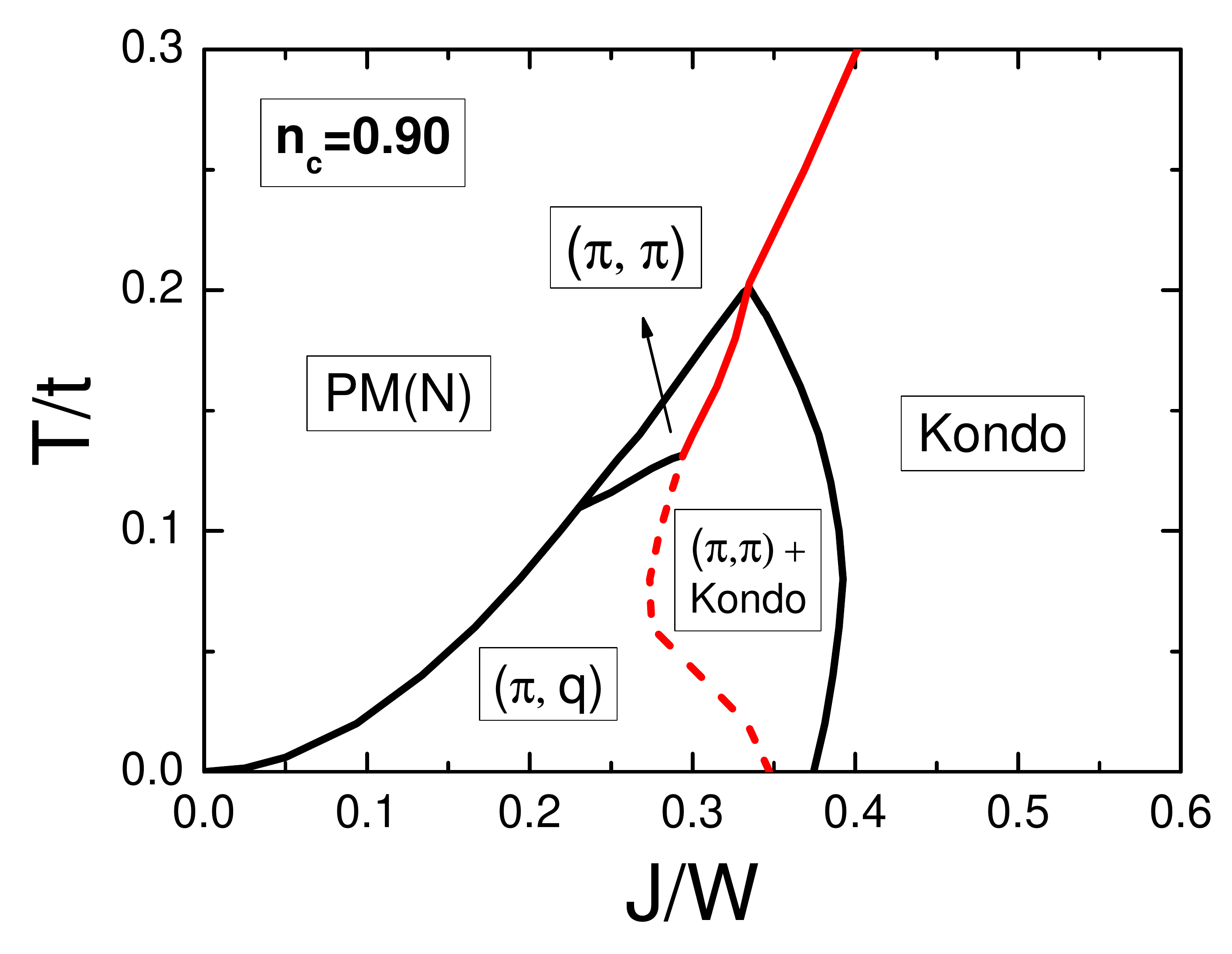}
\caption{(Color online) Same as Fig.\,\ref{fig:Doniach025}, but for $n_c=0.9$.}
\label{fig:Doniach09} % Fig 15
\end{figure}

For $n_c=0.9$, the phase diagram is shown in Fig.\,\ref{fig:Doniach09}.
Unlike the previous cases, by increasing the temperature one can now smoothly interpolate from $(\pi,q)$ to $(\pi,\pi)$, provided the value of $J/W$ lies entirely to the left of the first order line (red dashed curve in Fig.\,\ref{fig:Doniach09}).
By contrast, if one chooses, say $J/W=0.28$ as in Fig.\,\ref{fig:qvsT}(c), one obtains a reentrant behaviour for the AFM mode,  $\mathbf{Q}=(\pi,\pi)$.

%%%%%%%%%%%%%%%%%%%%%%   DISCUSSION   %%%%%%%%%%%%%%%%%%%%%%%%%%%%%%%%%%%%%%%%
\section{\label{sec5} Conclusions}

In conclusion, we have analysed the Kondo Lattice model on a square lattice, using a semi-classical--spin approach within a Hartree-Fock approximation.
This allowed us to probe the presence of spiral magnetic modes, which, for some ranges of parameters, turned out to be more stable than the ferromagnetic, antiferromagnetic, and paramagnetic modes, usually assumed to be the only ones at play.
The presence of spiral phases is in line with DMRG calculations for the one-dimensional case\cite{Garcia2004}, as well as for the two-leg ladder\cite{Xavier2004}, so that one may expect they are not an artifact of the approximations employed here.
Accordingly, we have obtained a ground state phase diagram in terms of the Kondo coupling, $J/W$ ($W$ is the bandwidth), and the conduction electron density, $n_c$.
As $n_c$ varies from 0 to 1 (half-filling), the weak- to moderate coupling region displays a variety of 
incommensurate phases [i.e., with continuously changing magnetic wavevectors, in the form $\mathbf{Q}=(\pi,q)$, $(q,0)$, or $(q,q)$], in which Kondo screening is absent. 
Recent dynamical mean-field theory (DMFT) calculations on the KLM away from half filling (see Ref.\,\onlinecite{R.Peters2015}) 
show incommensurate spin-density waves on both the small- and large Fermi surface regions; our results are in good agreement in the unscreened region.

We have also found that magnetic phases can coexist with some degree of screening, and that the change in magnetic wavevector with the Kondo coupling occurs even in the region of coexistence with the Kondo phase. 
Such a coexistence with incommensurate magnetic modes occurs in just a tiny region of the phase diagram, being related to the modulation of the triplet hybridisation term. 
It seems that a modulated hybridisation is an important ingredient for the stabilisation of magnetic modes other than $\mathbf{Q}=(0,0)$ or $\mathbf{Q}=(\pi,0)$ in the coexistence region. 
On the other hand, for sufficiently strong coupling, screening dominates and magnetism is suppressed.
%\color{blue}
While completing this work we became aware of Ref.\,\cite{HLi2015}, in which the decoupling of the Kondo term in singlet and triplet hybridisations is the same as ours, but, unlike our present framework, the magnetic wave vector $\mathbf{Q}$ was not left as a free parameter to be determined by minimisation of the free energy. 
Since Li \textit{et al.} \cite{HLi2015} set it as ($\pi, \pi$), they could not discuss spiral phases; instead, they considered the effects of an additional hopping term between next-nearest neighbours.
\color{black}  

We have also discussed the behaviour at finite temperatures in the form of `Doniach-like' phase diagrams $T\times J/W$, for fixed electronic densities. 
We have established that unscreened spiral magnetic phases can be found in the low-temperature and small-coupling portion of the phase diagram. 
Within each unscreened phase, the magnetic wavevector in general increases with temperature, until it reaches the first-order transition to the phase of screened magnetic order, when abrupt changes in $q$ may occur; interestingly, the coexisting magnetic mode is always commensurate, leading to antiferromagnetic, ferromagnetic, or striped phases.
As expected, at sufficiently high temperatures only paramagnetic phases survive, though they can be either unscreened or screened, depending on the magnitude of the Kondo coupling; however, 
as pointed out before\cite{Liu13}, the sharp transition between these two regimes is expected to become a crossover if fluctuations were taken into account beyond a mean-field treatment.  
The results presented here suggest that through a judicious choice of parameters, Kondo lattice systems may allow for temperature-driven switching between detectable magnetic modes. 
Notwithstanding the fact that the present results have been obtained for a square lattice, we have found  that the evolution of magnetic modes with the model parameters shares common trends with borocarbides family of materials, so that the Kondo lattice model should provide an adequate description of their non-superconducting properties.

%\color{blue}
We close with a cautionary remark. 
The prediction of ordered states at finite temperatures in two-dimen\-sional systems with continuous symmetry is certainly a drawback of mean-field approximations.  
Nonetheless, ground state features like the continuous variation of $\mathbf{Q}$ with the electronic density (as described in the simple picture above) should be present in both two- and three-dimensional systems; our results also broadly suggest how the temperature would influence the magnetic modes in three dimensions (or weakly-coupled two-dimensional layers). 
\color{black}

\begin{acknowledgments}

The authors are grateful to M.\,ElMassalami, M.\,A.\, Gusm\~ao, and T.\,Paiva for discussions, and to the Brazilian Agencies CAPES, CNPq, and FAPERJ for financial support.

\end{acknowledgments}

\appendix* %{MEAN-FIELD APPROXIMATION}
\section{Mean-field approximation}
\label{Ap}

The interaction term $\mathbf{S}_{i} \cdot \mathbf{s}^{c}_{i}$ in Eq.\,\eqref{hamiltonian} can be decomposed into several quartic operators, which are decoupled through a Hartree-Fock approximation (see, \textit{e.g.}, Ref.\,\onlinecite{Bruus}). 
In what follows, all mean values are taken into account to obtain the final mean-field Hamiltonian.

Using the definition of $ \mathbf{S}_{i}$ and $\mathbf{s}^{c}_{i}$ from Eqs.\,\eqref{Sf} and \eqref{Sc}, respectively, the axial component of the interaction term becomes
\begin{widetext}
\begin{align}
\label{Eq2_Ap}
\nonumber
S^{z}_{i} s^{z (c)}_{i}  
 & \approx \langle s^{z (c)}_{i} \rangle S^{z}_{i} + \langle S^{z}_{i} \rangle s^{z (c)}_{i} - \langle S^{z}_{i} \rangle \langle s^{z (c)}_{i} \rangle
+ \frac{1}{2} \langle V^{x}_{i c} \rangle  V^{x}_{i f} + \frac{1}{2} \langle V^{x}_{i f} \rangle  V^{x}_{i c} - \frac{1}{2} \langle V^{x}_{i f} \rangle \langle V^{x}_{i c} \rangle \\ \nonumber
& + \frac{1}{2} \langle V^{y}_{i c} \rangle  V^{y}_{i f} + \frac{1}{2} \langle V^{y}_{i f} \rangle  V^{y}_{i c} - \frac{1}{2} \langle V^{y}_{i f} \rangle \langle V^{y}_{i c} \rangle
- \frac{1}{2} \langle V^{z}_{i c} \rangle  V^{z}_{i f} - \frac{1}{2} \langle V^{z}_{i f} \rangle  V^{z}_{i c} + \frac{1}{2} \langle V^{z}_{i f} \rangle \langle V^{z}_{i c} \rangle \\ 
& - \frac{1}{2} \langle V^{0}_{i c} \rangle  V^{0}_{i f} - \frac{1}{2} \langle V^{0}_{i f} \rangle  V^{0}_{i c} + \frac{1}{2} \langle V^{0}_{i f} \rangle \langle V^{0}_{i c} \rangle ,
\end{align}
while the planar component can be written as
\begin{align}
%line 1
\label{Eq3_Ap}
\nonumber 
S^{x}_{i} s^{x (c)}_{i} + S^{y}_{i} s^{y (c)}_{i} & \approx
\langle s^{x (c)}_{i} \rangle S^{x}_{i} + \langle S^{x}_{i} \rangle s^{x (c)}_{i} - \langle S^{x}_{i} \rangle \langle s^{x (c)}_{i} \rangle 
+ \langle s^{y (c)}_{i} \rangle S^{y}_{i} + \langle S^{y}_{i} \rangle s^{y (c)}_{i} - \langle S^{y}_{i} \rangle \langle s^{y (c)}_{i} \rangle \\
& + \langle V^{z}_{i c} \rangle  V^{z}_{i f} +  \langle V^{z}_{i f} \rangle  V^{z}_{i c} -  \langle V^{z}_{i f} \rangle \langle V^{z}_{i c} \rangle
-  \langle V^{0}_{i c} \rangle  V^{0}_{i f} -  \langle V^{0}_{i f} \rangle  V^{0}_{i c} +  \langle V^{0}_{i f} \rangle \langle V^{0}_{i c} \rangle ,
\end{align}
with the definitions of $V^{\alpha}_{i \beta}$ ($\alpha=0,x,y,z$; $\beta=c, f$) given by Eqs.\,\eqref{singlet_hyb} and \eqref{triplet_hyb}.

Equations \eqref{Eq2_Ap} and \eqref{Eq3_Ap} then lead to
\begin{align}\label{Eq4_Ap}
\nonumber 
\mathbf{S}_{i} \cdot \mathbf{s}^{c}_{i} & \approx 
\langle \mathbf{s}^{c}_{i} \rangle \cdot \mathbf{S}_{i} + \langle \mathbf{S}_{i} \rangle \cdot \mathbf{s}^{c}_{i} - \langle \mathbf{S}_{i} \rangle \cdot \langle \mathbf{s}^{c}_{i} \rangle
+ \frac{1}{2} \langle \mathbf{V}_{i f} \rangle \cdot \mathbf{V}_{i c} + \frac{1}{2} \langle \mathbf{V}_{i c} \rangle \cdot \mathbf{V}_{i f} - \frac{1}{2} \langle \mathbf{V}_{i f} \rangle \cdot \langle \mathbf{V}_{i c} \rangle \\
& - \frac{3}{2} \langle V^{0}_{i f} \rangle V^{0}_{i c} - \frac{3}{2} \langle V^{0}_{i c} \rangle V^{0}_{i f} + \frac{3}{2} \langle V^{0}_{i f} \rangle \langle V^{0}_{i c} \rangle .
\end{align}
\end{widetext}
Then, substituting Eq.\,\eqref{Eq4_Ap} into the Hamiltonian, Eq.\,\eqref{hamiltonian}, leads to Eq.\,\eqref{hamil2}.

In addition, in order to fix the electronic density and the number of local magnetic moments, the terms $-\mu \big( \sum_{i \sigma} c^{\dagger}_{i\sigma} c^{\phantom{\dagger}}_{i\sigma} - N n_{c} \big)$ and $\epsilon_{f} \big( \sum_{i \sigma} f^{\dagger}_{i\sigma} f^{\phantom{\dagger}}_{i\sigma} - N n_{f} \big)$ must also be included in the Hamiltonian, Eq.\,\eqref{hamil2}, with $n_{f}=1$. 
These terms represent the constraints which are included as Lagrange multipliers, whose values of $\mu$ and $\epsilon_{f}$ are determined self-consistently.

For completeness, we recall that the mean values are expressed as
\begin{align}
& \langle \mathbf{S}_{i} \rangle = m^{0}_{f} \big( \cos \mathbf{Q} \!\cdot\! \mathbf{R_{i}}, \sin \mathbf{Q} \!\cdot\! \mathbf{R_{i}},0 \big) , \\
& \langle \mathbf{s}^{c}_{i} \rangle = -m^{0}_{c} \big( \cos \mathbf{Q}\!\cdot\! \mathbf{R_{i}}, \sin \mathbf{Q}\!\cdot\! \mathbf{R_{i}},0 \big), \\
& \langle V^{0}_{i c} \rangle = \langle {V^{0}_{i f}}^{\dagger} \rangle = -V_{0} 
\end{align}
and
\begin{equation}
\langle \mathbf{V}_{i c} \rangle  = \langle  \mathbf{V}^{\dagger}_{i f} \rangle = V_{0}' \big( \cos \left(\mathbf{Q}\!\cdot\! \mathbf{R}_{i}\right), \sin \left(\mathbf{Q}\!\cdot\! \mathbf{R}_{i}\right),0 \big).
\end{equation}

We now perform a discrete Fourier transform on the conduction electrons operators (and similarly for the \textit{f} electrons), defined as
\begin{equation}
c_{\mathbf{k} \sigma} = \frac{1}{\sqrt{N}}\sum_{i} \exp(i \mathbf{k}\!\cdot\!\mathbf{R}_{i}) c_{i\sigma} ,
\end{equation}
where \textit{N} is the number of lattice sites. 
Then, the Hamiltonian becomes
\begin{widetext}
\begin{align} \label{Eq5_Ap}
%line 1
\nonumber  \mathcal{H}_{MF} &= \sum_{\mathbf{k}} (\epsilon_{k} -\mu) c^{\dagger}_{\mathbf{k} \uparrow} c^{\phantom{\dagger}}_{\mathbf{k} \uparrow} 
%line 1
%\\ \nonumber & 
+ \sum_{\mathbf{k}} (\epsilon_{k+Q} -\mu) c^{\dagger}_{\mathbf{k}+\mathbf{Q} \downarrow} c^{\phantom{\dagger}}_{\mathbf{k}+\mathbf{Q} \downarrow}
%line 2
%\\ \nonumber & 
+ \frac{J m^{0}_{f}}{2} \sum_{\mathbf{k}} \big( c^{\dagger}_{\mathbf{k} \uparrow} c^{\phantom{\dagger}}_{\mathbf{k}+\mathbf{Q} \downarrow} + \mathrm{H.c.} \big)
%line 3
\\ \nonumber & 
- \frac{J m^{0}_{c}}{2} \sum_{\mathbf{k}} \big( f^{\dagger}_{\mathbf{k} \uparrow} f^{\phantom{\dagger}}_{\mathbf{k}+\mathbf{Q} \downarrow} + \mathrm{H.c.} \big)
+ \epsilon_{f} \sum_{\mathbf{k}} \big( f^{\dagger}_{\mathbf{k} \uparrow} f^{\phantom{\dagger}}_{\mathbf{k} \uparrow} + f^{\dagger}_{\mathbf{k} +\mathbf{Q} \downarrow} f^{\phantom{\dagger}}_{\mathbf{k} +\mathbf{Q} \downarrow} \big)
%line 4
\\ \nonumber &
+ \frac{J}{4} 3V_{0} \sum_{\mathbf{k}} \big( c^{\dagger}_{\mathbf{k} \uparrow} f^{\phantom{\dagger}}_{\mathbf{k} \uparrow} + c^{\dagger}_{\mathbf{k} +\mathbf{Q} \downarrow} f^{\phantom{\dagger}}_{\mathbf{k} +\mathbf{Q} \downarrow} + \mathrm{H.c.} \big) 
%line 5
%
%\\ \nonumber & 
+ \frac{J}{4} V_{0}' \sum_{\mathbf{k}} \big( c^{\dagger}_{\mathbf{k} \uparrow} f^{\phantom{\dagger}}_{\mathbf{k} + \mathbf{Q} \downarrow} + c^{\dagger}_{\mathbf{k} +\mathbf{Q} \downarrow} f^{\phantom{\dagger}}_{\mathbf{k} \uparrow} + \mathrm{H.c.} \big) 
\\ & + JNm^{0}_{f}m^{0}_{c} + \frac{3}{2}JNV_{0}^2 - \frac{1}{2}JNV_{0}'^2+ N n_{c} \mu - N n_{f} \epsilon_{f},
\end{align}
where $\epsilon_{k}= -2t\big[ \cos(k_{x}) + \cos(k_{y}) \big]$, while $\epsilon_{k + Q}= -2t\big[ \cos(k_{x} + q_{x}) + \cos(k_{y} + q_{y}) \big]$. Then, limiting ourselves to non-degenerate subspace $( \mathbf{k} \uparrow, \mathbf{k} + \mathbf{Q} \downarrow)$, where the base vectors are (in a Nambu spinor representation)
\begin{equation} \label{Nambu}
\Psi^{\dagger}_{\mathbf{k}} = \big( c^{\dagger}_{\mathbf{k} \uparrow}, c^{\dagger}_{\mathbf{k} + \mathbf{Q} \downarrow}, f^{\dagger}_{\mathbf{k} \uparrow}, f^{\dagger}_{\mathbf{k} + \mathbf{Q} \downarrow} \big),
\end{equation} 
the Hamiltonian is
\begin{equation}
\mathcal{H}_{MF} = \sum_{\mathbf{k}} \Psi^{\dagger}_{\mathbf{k}} \hat{H} (\mathbf{k}\!\uparrow, \mathbf{k}\!+\!\mathbf{Q}\!\downarrow) \Psi_{\mathbf{k}} + \mathrm{const.},
\end{equation}
where
\begin{equation}
\hat{H}(\mathbf{k}\!\uparrow, \mathbf{k}\!+\!\mathbf{Q}\!\downarrow)=
\left(
    {
    \begin{array}{cccccccc}
\epsilon_{k} - \mu & \frac{1}{2}J m^{0}_{f} & \frac{3}{4}J V_{0} & \frac{1}{4} J V_{0}' \\
\\
\frac{1}{2}J m^{0}_{f} & \epsilon_{k+Q} - \mu & \frac{1}{4} J V_{0}' & \frac{3}{4}J V_{0}  \\
\\
\frac{3}{4}J V_{0} & \frac{1}{4} J V_{0}' & \epsilon_{f} & -\frac{1}{2}J m^{0}_{c} \\
\\
\frac{1}{4} J V_{0}' & \frac{3}{4}J V_{0} & -\frac{1}{2}J m^{0}_{c} & \epsilon_{f} \\
    \end{array}
    }
  \right).
\label{eq:H4x4}  
\end{equation}
\end{widetext}
%\vskip 1cm

%\subsection{Ferromagnetism}

In some instances, the $4\times4$ Hamiltonian matrix \eqref{eq:H4x4} reduces to simpler $2\times2$ matrices, saving a significant amount of CPU time. 
One particularly interesting example is when the self-consistency process converges to a ferromagnetic state, that is, one with $\mathbf{Q}=(0,0)$. 
In order to determine the most general $2\times 2$ Hamiltonian matrix, we make use of the rotational symmetry, and take 
\begin{align}
&\langle \mathbf{S}_{i} \rangle = m^{0}_{f} \big( 0, 0, 1 \big), \label{ferro1} \\
&\langle \mathbf{s}^{c}_{i} \rangle = -m^{0}_{c} \big( 0, 0, 1 \big), \label{ferro2}
\end{align}
and
\begin{align}
&\langle \mathbf{V}_{i} \rangle = V_{0}' \big( 0, 0, 1 \big), \label{hyb2}
\end{align}
on the Hamiltonian of Eq.\,\eqref{hamil2}.

Then, taking Eqs.\,\eqref{ferro1} to \eqref{hyb2} into Eq.\,\eqref{hamil2}, the Hamiltonian becomes
\begin{widetext}
\begin{align}
%line 1
\nonumber \mathcal{H}_{MF} &= -t \sum_{\langle i,j \rangle, \sigma} \big( c^{\dagger}_{i \sigma} c^{\phantom{\dagger}}_{j \sigma} + H.c. \big)
%line 2
%\\ \nonumber & 
+ \frac{J m^{0}_{f}}{2} \sum_{i \sigma} \sigma c^{\dagger}_{i \sigma} c^{\phantom{\dagger}}_{i \sigma} - \frac{J m^{0}_{c}}{2} \sum_{i \sigma} \sigma f^{\dagger}_{i \sigma} f^{\phantom{\dagger}}_{i \sigma}
%line 3
%\\ \nonumber & 
+ \frac{J}{4}\big( 3V_{0} - V_{0}' \big) \sum_{i} \big( c^{\dagger}_{i \uparrow} f^{\phantom{\dagger}}_{i \uparrow} + f^{\dagger}_{i \uparrow} c^{\phantom{\dagger}}_{i \uparrow} \big)
\\  & 
+ \frac{J}{4}\big( 3V_{0} + V_{0}' \big) \sum_{i} \big( c^{\dagger}_{i \downarrow} f^{\phantom{\dagger}}_{i \downarrow} + f^{\dagger}_{i \downarrow} c^{\phantom{\dagger}}_{i \downarrow} \big)
%
%\\ \nonumber & 
+ \frac{3}{2}JNV_{0}^2 - \frac{1}{2}JNV_{0}'^2 + J N m^{0}_{f} m^{0}_{c}.
\end{align}
Fourier transforming the operators in the previous equation, and adding the constraint terms leads to
\begin{align}
\label{FT_hamil4x4}
%line 1
\nonumber & \mathcal{H}_{MF} = \sum_{ \mathbf{k}, \sigma} \big( \epsilon_{k} - \mu + \frac{\sigma J m^{0}_{f}}{2} \big) c^{\dagger}_{\mathbf{k} \sigma} c^{\phantom{\dagger}}_{\mathbf{k} \sigma}
%line 2
%\\ \nonumber & 
+ \sum_{ \mathbf{k}, \sigma} \big( \epsilon_{f} - \frac{\sigma J m^{0}_{c}}{2} \big) f^{\dagger}_{\mathbf{k} \sigma} f^{\phantom{\dagger}}_{\mathbf{k} \sigma}
%line 3
%\\ \nonumber & 
+ \frac{J}{4} \sum_{\mathbf{k}, \sigma} \big( 3V_{0} - \sigma V_{0}' \big) \big( c^{\dagger}_{\mathbf{k} \sigma} f^{\phantom{\dagger}}_{\mathbf{k} \sigma} + f^{\dagger}_{\mathbf{k} \sigma} c^{\phantom{\dagger}}_{\mathbf{k} \sigma} \big)
\\  & + \frac{3}{2}JNV_{0}^2 - \frac{1}{2}JNV_{0}'^2 + J N m^{0}_{f} m^{0}_{c}
%
%\\ \nonumber & 
+ \mu n_{c}N - \epsilon_{f} n_{f} N.
\end{align}
Using a Nambu spinor representation $
\Psi^{\dagger}_{\mathbf{k} \sigma} = \big( c^{\dagger}_{\mathbf{k} \sigma}, f^{\dagger}_{\mathbf{k} \sigma} \big)$, it becomes
\begin{equation}
\label{FT_hamil2x2}
\mathcal{H}_{MF} = \sum_{\mathbf{k} \sigma} \Psi^{\dagger}_{\mathbf{k} \sigma} \hat{H}(\mathbf{k} \sigma) \Psi_{\mathbf{k} \sigma} + const.,
\end{equation}
where
\begin{equation}
\hat{H}(\mathbf{k} \sigma)=
\left(
    {
    \begin{array}{cccc}
\tilde{\epsilon}_{k} + \frac{\sigma J m^{0}_{f}}{2} & \frac{J}{4}\big( 3V_{0} - \sigma V_{0}'\big)
\\
\\
\frac{J}{4}\big( 3V_{0} - \sigma V_{0}'\big) & \epsilon_{f} - \frac{\sigma J m^{0}_{c}}{2}
    \end{array}
    }
  \right),
\end{equation}
which provides the eigenvalues
\begin{align}
\label{eq:energies_FM}
E^{\pm}_{\mathbf{k} \sigma} = \frac{1}{2} \bigg[ \tilde{\epsilon}_{k} + \epsilon_{f} + \frac{\sigma J}{2} (m^{0}_{f} - m^{0}_{c}) \bigg] %\\\nonumber & 
\pm \frac{1}{2} \sqrt{\bigg[ \tilde{\epsilon}_{k} - \epsilon_{f} + \frac{\sigma J}{2} (m^{0}_{f} + m^{0}_{c}) \bigg]^2 + \frac{J^2}{4}\big( 3V_{0} - \sigma V_{0}' \big)^2},
\end{align}
which $\tilde{\epsilon}_{k} = \epsilon_{k} - \mu$.
\end{widetext}

It is interesting to note that the spectra of the $4\times4$ and of the two $2\times2$ representations of the mean-field Hamiltonian are equivalent when $\mathbf{Q}=(0,0)$, irrespective of $V_{0}$ and $V_{0}'$ vanishing or not. 

\bibliography{bib_KLM}

\end{document}